\documentclass[pra, twocolumn, superscriptaddress]{revtex4-2}
\usepackage{graphicx}
\usepackage{amssymb}
\usepackage{amsfonts}
\usepackage{amsmath}
\usepackage{lmodern}
\usepackage{mathrsfs}
\usepackage{mathtools, slashed}
\usepackage{mathdots}
\usepackage{collref}
\usepackage[colorlinks=true, citecolor=blue, urlcolor=blue]{hyperref}
\usepackage{tikz}
\usepackage{bm}
\usepackage[markup=underlined]{changes}
\makeatletter
\AddToHook{cmd/added/before}{\def\Changes@AuthorColor{purple}}
\AddToHook{cmd/deleted/before}{\def\Changes@AuthorColor{green}}
\AddToHook{cmd/replaced/before}{\def\Changes@AuthorColor{red}}
\makeatother
\usepackage{todonotes}
\setcommentmarkup{\todo[color={pink}, size=\scriptsize]{#3: #1}}

\usetikzlibrary{shapes.geometric}

\begin{document}
\title{Metastability in the Dissipative Quantum Rabi Model}

\author{Da-Wu Xiao}
\email{dawuxiao@gmail.com}
\author{Chong Chen}
\email{chongchenn@gmail.com}
\thanks{\\These authors contributed equally to this work.}
\affiliation{Department of Physics, Chinese University of Hong Kong, Shatin, New Territories, Hong Kong, China}%
\affiliation{State Key Laboratory of  Quantum Information Technologies and Materials, Chinese University of Hong Kong, Shatin, New Territories, Hong Kong, China}%
\affiliation{New Cornerstone Science Laboratory, Chinese University of Hong Kong, Shatin, New Territories, Hong Kong, China}%
\date{\today}

\begin{abstract}
The dissipative quantum Rabi model exhibits rich non-equilibrium physics, including a dissipative phase transition from the normal phase to the superradiant phase. In this work, we investigate the stability of the superradiant phase in the presence of a weak spin relaxation. We find that even a weak spin relaxation can render the superradiant phase a superradiant metastable phase, in which symmetry-breaking states are stable only for a finite time.  This arises because each spin jump induced by relaxation applies as a strong perturbation to the system, potentially driving the system from a symmetry-breaking state to the symmetry-preserving saddle point with finite probability, before it eventually relaxes back to a symmetry-breaking state. Such dynamical processes lead to a finite lifetime of the symmetry-breaking states and restore the symmetry in the steady state. To substantiate these results, we combine mean-field and cumulant expansion analyses with exact numerical simulations. The lifetimes of the symmetry-breaking states are analyzed in finite-size systems and the conclusions are extrapolated to the thermodynamic limit via finite-size scaling. Our findings establish the metastable nature of the symmetry-breaking states in the dissipative quantum Rabi model and reveal the complexity of the dissipative phase transition beyond their equilibrium counterpart. The mechanisms uncovered here can be generalized to a broad class of open quantum systems, highlighting fundamental distinctions between equilibrium phase transitions and steady-state phase transitions.
\end{abstract}
\maketitle

\section{Introduction}
The quantum Rabi model (QRM), describing a two-level atom (or a spin $1/2$) coupled to a single bosonic mode, is one of the simplest yet most profound models in quantum physics~\cite{Rabi1936}. It has inspired extensive fundamental studies that have reshaped our understanding of light matter interactions~\cite{Brune1996,Zagoskin2008,Ashhab2010,Lo2015}, strong-coupling physics~\cite{Niemczyk2010,Yoshihara2016,FornDiaz2019}, integrability~\cite{Braak2011}, and quantum phase transitions~\cite{Sachdev2011}. Notably, recent studies have revealed that the QRM exhibits a superradiant phase transition in the limit where the frequency ratio between the two-level atom and the bosonic mode tends to infinity~\cite{Ashhab2013,Bakemeier2012,Hwang2015, Liu2017,Zhu2020,Felicetti2020}. Moreover, subsequent studies have demonstrated that the QRM can also undergoes a dissipative phase transition when the bosonic mode is subject to relaxation, providing a versatile framework for exploring nonequilibrium critical phenomena~\cite{Hwang2018,Lyu2024,Li2024,Ilias2022, Yang2023}. 
These findings broaden our understanding of phase transitions by showing their occurrence even in systems comprising only a few interacting components and also highlight promising applications in quantum critical sensing~\cite{Garbe2020, Chu2021, Beaulieu2025}.

Dissipative phase transition manifest as abrupt changes in the steady state of an open quantum system and exhibit strong analogies to equilibrium phase transitions through the correspondence between the Liouvillian and Hamiltonian~\cite{Kessler2012,Minganti2018}. However, due to their nonequilibrium nature, dissipative phase transitions exhibit richer and more complex phenomena than their equilibrium counterparts, including  bistability~\cite{MendozaArenas2016,Sibalic2016,Melo2016,Alaeian2022}, intermittency~\cite{Lee2012,Ates2012}, multimodality~\cite{Malossi2014,Letscher2017}, metastability~\cite{ Macieszczak2016, Rose2016,LeBoite2017, Macieszczak2021, Jin2024}, multicriticality~\cite{Soriente2018, Ferri2021}, dynamical phase transitions~\cite{Hedges2009, Lesanovsky2013}, and phase transition without symmetry breaking~\cite{Hannukainen2018, Minganti2021}. Moreover, the stability of nonequilibrium phases under perturbations remains an open question~\cite{Letscher2017,Rakovszky2024}.
In particular, metastable states can emerge in nonequilibrium systems due to the perturbations from both classical and quantum fluctuations,  which undermine the stability of the non-equilibrium phase.
Such states are characterized by a separation of timescales in the system dynamics~\cite{ Macieszczak2016}, with one timescale being much longer than all others, a feature commonly observed in finite-size systems undergoing first-order phase transition.  The dissipative QRM provides an insightful starting point to explore this issue.

Here we investigate the stability of the superradiant phase in the dissipative QRM under a weak spin relaxation. The key idea is that, since only a single spin is involved in the QRM, spin quantum jumps induced by relaxation act as strong perturbations to the system and thereby undermine the stability of different phases in the dissipative QRM, particularly the symmetry-breaking phase. On the other hand, the phase diagram of the dissipative QRM is usually determined using mean-field theory. However, mean-field results become invalid when large classical or quantum fluctuations are present. For instance, multiple stationary states predicted by mean-field theory can be ruled out by more sophisticated approaches, such as variational methods~\cite{Weimer2015,Weimer2015a} and infinite tensor network simulations~\cite{Gangat2017}. Therefore, it is crucial to study the stability of the superradiant phase in the dissipative QRM beyond mean-field theory. Yet, this problem has rarely been addressed in previous works.

In this paper we show that the superradiant phase in the dissipative QRM is generally metastable under a weak spin relaxation. This conclusion is supported by analyses based on the cumulant expansion calculations, exact Husimi $Q$ representation of the steady state, exact quantum trajectory simulations, and Liouvillian gap analysis. In particular, finite-size effects are excluded through finite-size-scaling analysis. These results demonstrate that even a weak spin relaxation transforms the superradiant phase into a superradiant metastable phase, where symmetry-breaking states only have a finite lifetime. Our work provides a comprehensive analytical and numerical framework for investigating dissipative phase transitions in open quantum systems. 

This paper is organized as follows. In Sec.~\ref{sec:DPT} we introduce the dissipative QRM and derive its phase diagram using mean-field theory, linear stability analysis, and the cumulant expansion method.
Motivated by the divergence of cumulant expansion in the superradiant phase, we analyze its stability under strong perturbations, revealing that this phase is metastable. 
In Sec.~\ref{sec:Exact} we employ exact numerical simulations to uncover the mechanism of the metastability. The Husimi $Q$ representation of the steady state reveals that spin relaxation opens transition channels between symmetry-breaking states, thereby leading to a finite lifetime of the symmetry-breaking states—an explicit signature of metastability. These findings are further substantiated by quantum trajectory simulations.
In Sec.~\ref{sec:Lifetime} we use the Liouvillian gap to quantify the lifetime of the symmetry-breaking states.  A finite-size-scaling analysis demonstrates that this lifetime remains finite in the thermodynamic limit, thereby confirming the metastability of the superradiant phase in the presence of a weak spin relaxation. We present a discussion and conclusion in Sec.~\ref{sec:Discussion} and summarize in Sect.~\ref{sec:Conclusion}.

\section{Dissipative phase transition and stability analysis}\label{sec:DPT}
\begin{figure}
    \centering
    \includegraphics[width=1.0\columnwidth]{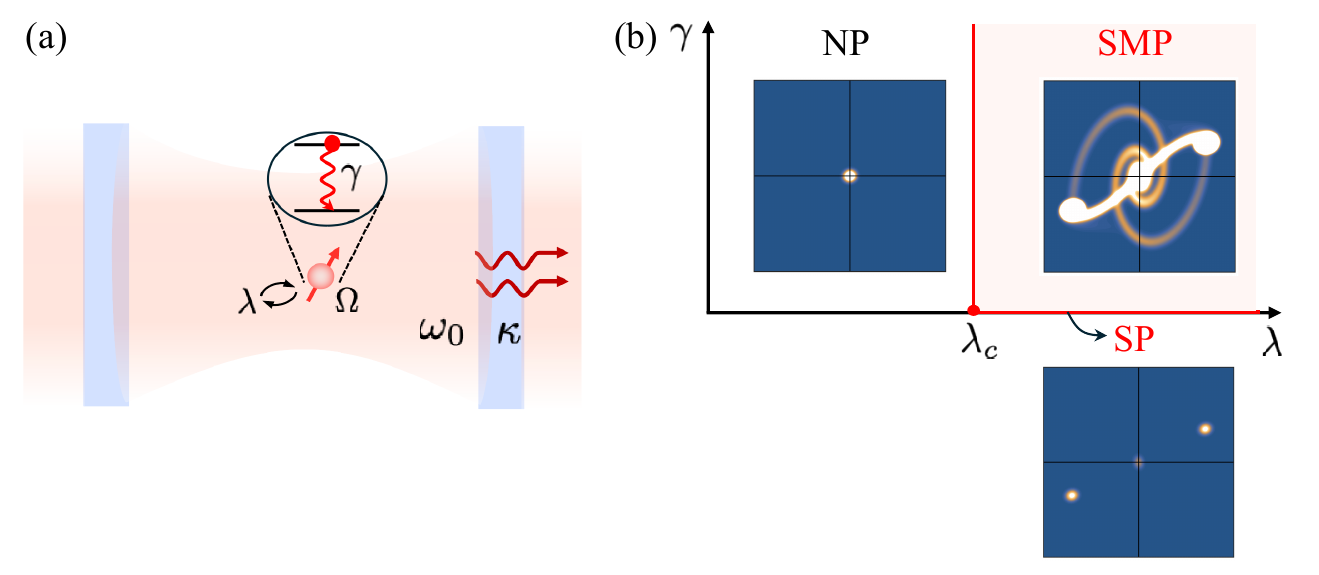}
    \caption{(a) Schematic diagram of the dissipative quantum Rabi model under a weak spin relaxation and (b) its corresponding phase diagram. Here, NP denotes the normal phase, SP denotes the superradiant phase,  and SMP denotes the superradiant metastable phase. When $\gamma=0$, the system undergoes a phase transition from the NP to the SP at the critical point $\lambda_c$. When $\gamma\neq0$,  this critical point extends into a critical line, and the SP transforms into the SMP. The inset shows the typical Husimi $Q$ representation of the bosonic mode in different phases. }
    \label{fig:Scheme}
\end{figure}

The quantum Rabi model describes a two-level atom (or a spin-1/2) coupled to a single bosonic mode~\cite{Braak2011,Hwang2015}. 
The Hamiltonian reads (with $\hbar =1$)
\begin{equation}
    \hat{H}=\omega_{0}\hat{a}^{\dagger}\hat{a}+\frac{\Omega}{2}\hat{\sigma}_{z}+\frac{\lambda}{\sqrt{2}}\left(\hat{a}^{\dagger}+\hat{a}\right)\hat{\sigma}_{x},
\end{equation}
where $\hat{a}$ ($\hat{a}^{\dagger}$) is the annihilation (creation) operator of the bosonic mode with frequency $\omega_0$, $\hat{\sigma}_{x,y,z}$ are the Pauli operators of the spin with energy splitting $\Omega$, and $\lambda$ denotes the coupling strength between the spin and the bosonic mode. The QRM possesses parity symmetry, denoted by $[\hat{\mathcal{P}},\hat{H}]=0$ with $\hat{\mathcal{P}}\equiv e^{i\pi(\hat{a}^{\dagger}\hat{a}+\frac{1}{2}\hat{\sigma}_{z})}$ the parity operator. By introducing the frequency ratio $\Omega/\omega_0$ as a scaling parameter in the QRM, Hwang {\it et al.} revealed that the system exhibits a second-order quantum phase transition at the critical coupling strength $\lambda_c=\sqrt{\omega_{0}\Omega/2}$ in the limit $\Omega/\omega_0\rightarrow \infty$. As the coupling strength $\lambda$ changes, the ground state undergoes a transition from a parity-preserving normal phase to a parity-breaking superradiant phase~\cite{Bakemeier2012,Hwang2015,Liu2017}. Taking into account relaxation processes from both the bosonic mode and the spin due to interactions with external environments, the dissipative QRM is described by the master equation
\begin{equation}\label{eq:DRM}
    \frac{d\rho}{dt}=-i[\hat{H},\rho]+\mathcal{D}_{\hat{a}}\left[\rho\right]+\mathcal{D}_{\hat{\sigma}_{-}}\left[\rho\right],
\end{equation}
where $\rho$ denotes the state of the system and $\mathcal{D}_{\hat{a}}\left[\rho \right]\equiv 2\kappa\hat{a}\rho \hat{a}^{\dagger}-\kappa \hat{a}^{\dagger}\hat{a}\rho-\kappa \rho\hat{a}^{\dagger}\hat{a}$ and  $\mathcal{D}_{\hat{\sigma}_{-}}\left[\rho\right]\equiv 2\gamma \hat{\sigma}_{-}\rho\hat{\sigma}_{+}-\gamma \hat{\sigma}_{+}\hat{\sigma}_{-}\rho-\gamma \rho\hat{\sigma}_{+}\hat{\sigma}_{-} $ are the dissipators of the bosonic mode and the spin, with relaxation rates $\kappa$ and $\gamma$, respectively. The schematic diagram of this model is shown in Fig.~\ref{fig:Scheme}(a).  Notably, parity is still preserved under these dissipative processes, as $\hat{\mathcal{P}} (\mathcal{D}_{\hat{a}/\hat{\sigma}_{-}}[\rho] )\hat{\mathcal{P}}^{\dagger}=\mathcal{D}_{\hat{a}/\hat{\sigma}_{-}}[\hat{\mathcal{P}}\rho\hat{\mathcal{P}}^{\dagger}]$. In the absence of spin relaxation, i.e., $\gamma=0$, Hwang {\it et al.} also showed that the system still exhibits a dissipative phase transition at the critical coupling strength $\lambda_{c}=\sqrt{\frac{\omega_{0}\Omega}{2}}\sqrt{1+\frac{\kappa^{2}}{\omega_{0}^{2}}}$, where the steady state changes from the normal phase ($\lambda<\lambda_c$)  to the superradiant phase ($\lambda>\lambda_c$) ~\cite{Hwang2018}. Beyond the $\gamma=0$ regime, the phase diagram and the stability of different phases have received little attention.

\subsection{Mean-field theory}
In the presence of a weak spin relaxation ($\gamma \ll \omega_0$), the phase diagram can be analyzed with mean-field theory. The evolution equations of observables $\left\langle \hat{x}\right\rangle \equiv \langle\frac{ \hat{a}^{\dagger}+ \hat{a} }{\sqrt{2}}\rangle$, $\langle \hat{p}\rangle \equiv \langle \frac{\hat{a} - \hat{a}^{\dagger}}{\sqrt{2}i}\rangle$, $\langle \hat{\sigma}_{x}\rangle$, $\langle \hat{\sigma}_{y}\rangle$, and $\langle \hat{\sigma}_{z}\rangle$ are derived from Eq. \eqref{eq:DRM} by neglecting the quantum fluctuations, i.e. high-order correlations such as $\left\langle \hat{x}\hat{\sigma}_{y}\right\rangle-\langle \hat{x} \rangle \langle \hat{\sigma}_{y}\rangle$. The resulting mean-field equations are
\begin{align}\label{eq:Mean-field}
    &\frac{d\left\langle \hat{x}\right\rangle }{dt}=-\kappa\langle \hat{x}\rangle+\omega_{0}\left\langle \hat{p}\right\rangle, \\
    &\frac{d\left\langle \hat{p}\right\rangle }{dt}=-\kappa\left\langle \hat{p}\right\rangle -\omega_{0}\left\langle \hat{x}\right\rangle  -\lambda\left\langle \hat{\sigma}_{x}\right\rangle ,\\
    &\frac{d\left\langle \hat{\sigma}_{x}\right\rangle }{dt}=-\gamma\left\langle \hat{\sigma}_{x}\right\rangle-\Omega\left\langle \hat{\sigma}_{y}\right\rangle ,\\
    &\frac{d\left\langle \hat{\sigma}_{y}\right\rangle }{dt}=-\gamma\left\langle \hat{\sigma}_{y}\right\rangle+\Omega\left\langle \hat{\sigma}_{x}\right\rangle -2\lambda\left\langle \hat{x}\right\rangle \left\langle \hat{\sigma}_{z}\right\rangle  ,\\
    &\frac{d\left\langle \hat{\sigma}_{z}\right\rangle }{dt}=-2\gamma\left\langle \hat{\sigma}_{z}\right\rangle +2\lambda\left\langle \hat{x}\right\rangle \left\langle \hat{\sigma}_{y}\right\rangle-2\gamma.
\end{align}
Solving the above equations yields three steady-state solutions: one parity-preserving and a pair of parity-breaking ones 
\begin{equation} \label{eq:MFS-NP}
    \langle \hat{\sigma}_{z}\rangle_{s} =-1, \langle \hat{\sigma}_{x}\rangle_{s} =\langle\hat{\sigma}_{y}\rangle_{s} =\langle \hat{x}\rangle_{s} =\langle \hat{p}\rangle_{s} =0;
\end{equation}
and
\begin{equation}    \label{eq:MFS-SMP}
\begin{aligned}
    &\langle \hat{\sigma}_{z}\rangle_{s} =-\frac{\lambda_{c}^{2}}{\lambda^{2}},\langle \hat{\sigma}_{x}\rangle_{s} =\pm\sqrt{\frac{2}{1+\frac{\gamma^{2}}{\Omega^2}}\frac{\lambda_{c}^{2}}{\lambda^{2}}\left(1-\frac{\lambda_{c}^{2}}{\lambda^{2}}\right)}, \\
    & \frac{\langle \hat{\sigma}_{y}\rangle_{s}}{\langle \hat{\sigma}_x \rangle_{s}} =-\frac{\gamma}{\Omega},
    \frac{\langle \hat{x}\rangle_{s}}{\langle \hat{\sigma}_{x}\rangle_{s}} =-\frac{1}{1+\frac{\kappa^{2}}{\omega^2_0}}  \frac{\lambda}{\omega_0},\frac{\langle \hat{p}\rangle_{s}}{\langle \hat{x} \rangle_{s}} =\frac{\kappa}{\omega_0} 
\end{aligned}
\end{equation}
with $\lambda_{c}=\sqrt{\frac{\omega_{0}\Omega}{2}\left(1+\frac{\gamma^{2}}{\Omega^{2}}\right)\left(1+\frac{\kappa^{2}}{\omega_{0}^{2}}\right)}$. When $\lambda<\lambda_{c}$, only the parity-preserving solution given in Eq.~\eqref{eq:MFS-NP} is physical, as parity-breaking solutions in Eq.~\eqref{eq:MFS-SMP}  yield the non-physical result $\langle \hat{\sigma}_{z} \rangle<-1$. The system retains parity in this solution and thus is in the normal phase (NP). When $\lambda>\lambda_{c}$, all three steady-state solutions are physical. However, later stability analysis shows that the parity-preserving solution is unstable and the pair of parity-breaking solutions are metastable. The system is thus in the superradiant metastable phase (SMP).  The corresponding phase diagram is shown in Fig.~\ref{fig:Scheme}(b).

\subsection{Linear stability analysis}
To analyze the stability of the steady-state solutions, we linearize the mean-field equations by expanding each observable as $\left\langle \hat{o}\right\rangle =\left\langle \hat{o}\right\rangle _{s}+\delta\langle \hat{o}\rangle$, where $\delta \langle \hat{o}\rangle$ represents a perturbation around the steady-state solution $\left\langle \hat{o}\right\rangle _{s}$. Introducing the perturbation vector $\delta\vec{v}=\left(\delta \langle \hat{x} \rangle ,\delta \langle \hat{p}\rangle ,\delta \langle \hat{\sigma}_{x}\rangle ,\delta\langle \hat{\sigma}_{y}\rangle ,\delta \langle \hat{\sigma}_{z}\rangle \right)$ and neglecting nonlinear terms, we obtain a linearized equation $\frac{d\delta\vec{v}}{dt}=\mathbb{M}\delta\vec{v}$ of perturbations, where the coefficient matrix $\mathbb{M}$ reads
\begin{equation}
    \mathbb{M}=\left(\begin{array}{ccccc}
    -\kappa & \omega_{0} & 0 & 0 & 0\\
    -\omega_{0} & -\kappa & -\lambda & 0 & 0\\
    0 & 0 & -\gamma & -\Omega & 0\\
    -2\lambda \langle \hat{\sigma}_{z}\rangle _{s} & 0 & \Omega & -\gamma & -2\lambda\langle \hat{x}\rangle _{s}\\
    -2\lambda \langle \hat{\sigma}_{y}\rangle _{s} & 0 & 0 & 2 \lambda \left\langle \hat{x}\right\rangle _{s} & -2\gamma
    \end{array}\right).
\end{equation}
The stability of a steady-state solution is determined by the eigenvalues of $\mathbb{M}$. 
If all eigenvalues have negative real parts, $\delta\vec{v}$ decays to zero for any initial perturbation, implying that the steady-state solution is stable. In contrast, if any eigenvalue has a positive real part, perturbations along the direction of the corresponding eigenvector grow exponentially, driving the system away from the steady state and thus rendering it unstable.
Specifically, in the NP, we find that all eigenvalues corresponding to the parity-preserving solution have negative real parts, indicating that this solution is stable. In the SMP, however, the parity-preserving solution becomes a saddle point (unstable) as one eigenvalue has a positive real part. By contrast, all eigenvalues corresponding to the parity-breaking solutions have negative real parts, indicating that these solutions are stable.

\begin{figure}
    \centering
    \includegraphics[width=0.9\linewidth]{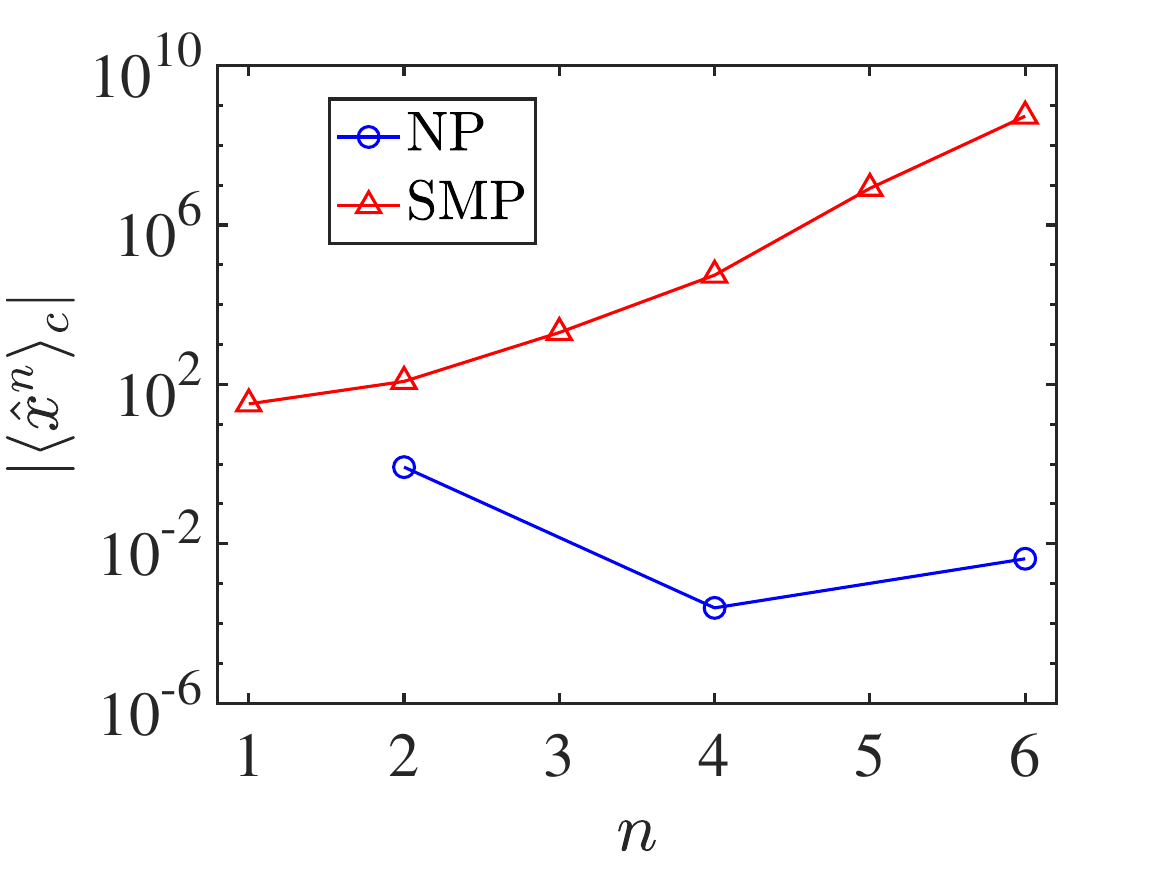}
    \caption{High-order cumulants $|\langle \hat{x}^n\rangle_c|$ obtained from solving the steady-state equations of cumulants truncated at sixth order. Odd order cumulants vanish in the NP as $\langle \hat{x}^n\rangle_c=0$ holds due to the parity symmetry. The parameters are $\omega_0=1.0$, $\kappa/\omega_0=0.5$, $\gamma/\omega_0=0.05$, and $\Omega/\omega_0=1200$. For the NP case, $\lambda=0.6\sqrt{{\omega_0 \Omega}/{2}}$, and for the SMP case, $\lambda=1.4\sqrt{{\omega_0 \Omega}/{2}}$.}
    \label{fig:cumulant}
\end{figure}

The above linear stability analysis is applicable only when quantum fluctuations are small. However, in many open quantum systems, quantum fluctuations can be significant and may drive the system away from the steady state predicted by mean-field theory.
To account for this effect, we extend the mean-field equations by incorporating quantum fluctuations through higher-order cumulants~\cite{Barquilla2020, Leymann2014,plankensteiner2022}. Formally, cumulants are defined via the generating function $\langle e^{\sum_{l} \eta_l \hat{o}_l}\rangle$ as
\begin{equation} \label{eq:cumulant}
    \langle \hat{o}^{n_1}_{i_1} \hat{o}^{n_2}_{i_2} \cdots \rangle_c\equiv \frac{d^{n_1}}{d \eta^{n_1}_{i_1}} \frac{d^{n_{2}}}{d\eta^{n_2}_{i_2}} \cdots \left. \ln \langle e^{\sum_{l} \eta_l \hat{o}_l}\rangle\right|_{\vec{\eta}=\vec{0}},
\end{equation}
where  $n \equiv n_1+n_2+\cdots$ denotes the order of the cumulant and $\hat{o}_i$ can be any operator of the bosonic mode or the spin~\cite{Kubo1962}. Higher-order cumulants ($n\geq2$) capture deviations from mean-field theory due to quantum fluctuations. From Eq. \eqref{eq:DRM} one can derive a hierarchy of equations for these cumulants, where the lower-order cumulants couple with the higher-order ones.  To make the equations tractable, we truncate the hierarchy at a given order by neglecting all higher-order cumulants (see Appendix~\ref{sec:CEM-app} for details).

In Fig.~\ref{fig:cumulant} we present the steady state value of $\langle \hat{x}^n\rangle_c$ obtained by solving the steady-state equations of cumulants up to sixth order around the mean-field solution after neglecting seventh and higher-order cumulants.    In the NP, higher-order cumulants converge rapidly, confirming that higher-order fluctuations are negligible. This indicates that the linear stability analysis is reliable in this phase. In contrast,  high-order cumulants in the parity-breaking solution diverge with the expansion order $n$ in the SMP. Such divergence indicates the presence of strong quantum fluctuations in the SMP, rendering the linear stability analysis unreliable in this regime. Our exact simulations further demonstrate that these quantum fluctuations, arising from both the intrinsic quantum fluctuation (which vanishes in the thermodynamic limit) and the spin relaxation process (which persists as long as $\gamma\neq0$), lead to the metastability of the parity-breaking solutions.

\subsection{Stability under strong perturbations}\label{subsec:StablityUP}

\begin{figure}
    \centering
    \includegraphics[width=1.0\linewidth]{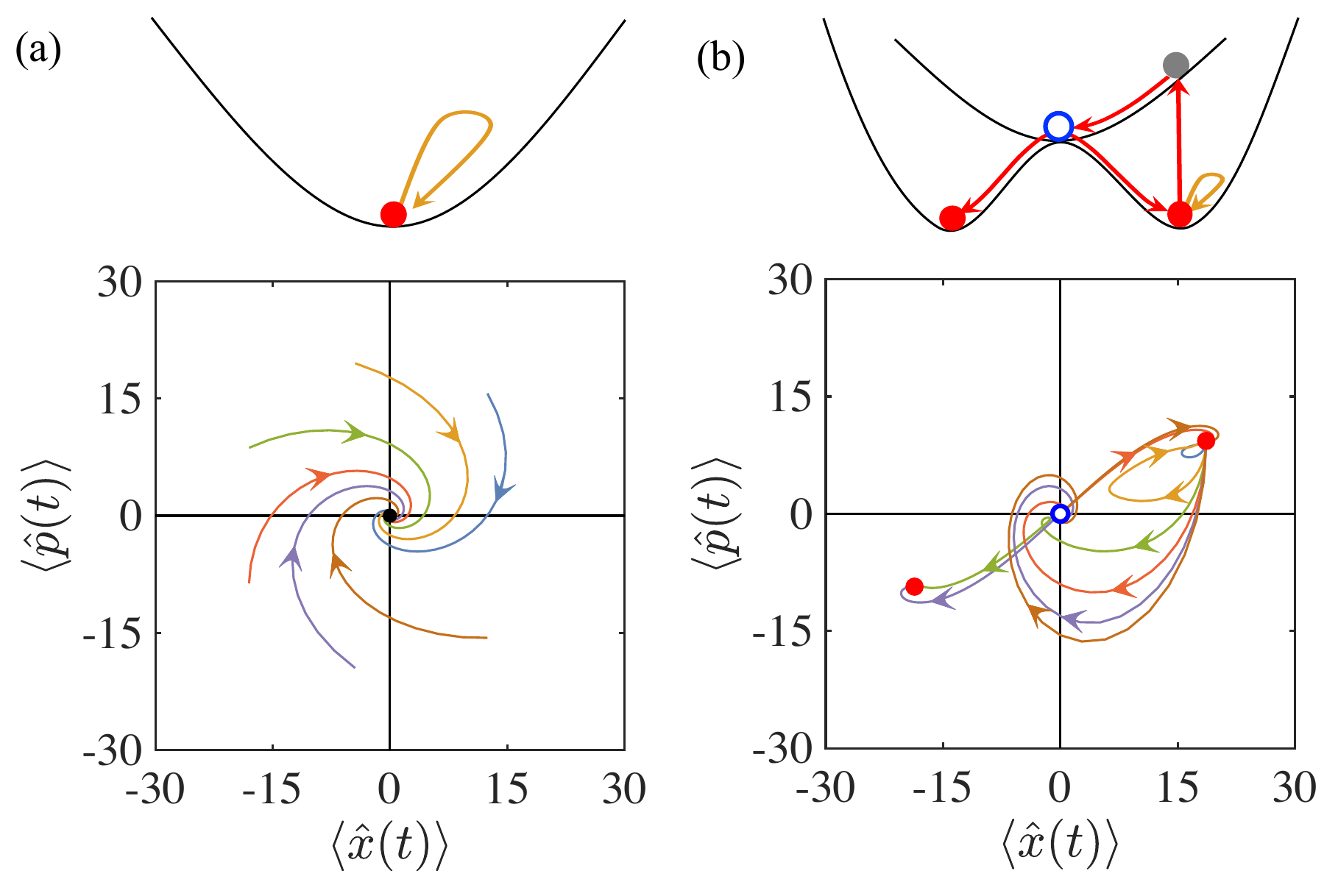}
    \caption{Evolution of the steady-state solution in the (a) NP and (b) SMP under strong perturbations within the mean-field framework. Phenomenological interpretations are shown in the top panels.  For the NP, we consider a quench of the parity-preserving solution with $\langle \hat{x}\rangle=r\cos 2\theta$ and $\langle \hat{p}\rangle= r \sin 2\theta$. For the SMP, we consider a quench of one parity-breaking solution with $\langle \hat{\sigma}_x\rangle=\cos \theta \langle \hat{\sigma}_x \rangle_{s}+\sin \theta \langle \hat{\sigma}_z \rangle_{s}$ and $\langle \hat{\sigma}_z\rangle=\cos \theta \langle \hat{\sigma}_z \rangle_{s}-\sin \theta \langle \hat{\sigma}_x \rangle_{s}$. We set $r=10$ and $\theta=\frac{\pi}{7},\frac{2\pi}{7}, \cdots, \frac{6\pi}{7}$. The other parameters are the same as those in Fig. \ref{fig:cumulant}.}
    \label{fig:StrongPerturb}
\end{figure}

To account for possible strong quantum fluctuations, we examine the stability of the steady state under strong perturbations within the mean-field framework. The results are shown in Fig.~\ref{fig:StrongPerturb}. 

In the NP ($\lambda < \lambda_c$), we consider a quench of the parity-preserving solution by setting $\langle \hat{x}\rangle=r \cos 2\theta$ and $\langle \hat{p}\rangle=r \sin 2\theta$.  In this phase, since the system has only one steady-state solution, the system under any perturbation flows back to the parity-preserving solution. A phenomenological description of such a process is shown in the top panel of Fig.~\ref{fig:StrongPerturb}(a).  Numerical results in the bottom panel confirm this conclusion.  In the SMP ($\lambda>\lambda_{c}$), the quench of the parity-breaking solution is set to be  $\langle \hat{\sigma}_x\rangle=\cos \theta \langle \hat{\sigma}_x \rangle_{s}+\sin \theta \langle \hat{\sigma}_z \rangle_{s}$ and $\langle \hat{\sigma}_z\rangle=\cos \theta \langle \hat{\sigma}_z \rangle_{s}-\sin \theta \langle \hat{\sigma}_x \rangle_{s}$, which is designed to mimic the spin flip due to spin relaxation. In this regime, the system possesses three steady-state solutions: a pair of parity-breaking solutions (which are stable under small perturbations) and one unstable parity-preserving solution (which acts as a saddle point).  A sufficiently strong perturbation, in principle, can drive the system from its local stable state to the saddle point, from which it eventually relaxes back into one of the parity-breaking states. A phenomenological description of such a process is shown in the top panel of Fig.~\ref{fig:StrongPerturb}(b). The numerical results, shown in bottom panel, demonstrate this behavior. Starting from a parity-breaking solution, the system remains stable under weak perturbations (small $\theta$). However, under strong perturbations (large $\theta$), it is driven away from its local stable state, flows toward the saddle point,  and eventually settles into one of the parity-breaking solutions. Consequently, the parity-breaking solutions are metastable under strong perturbations. 

From the cumulant analysis, we have shown that strong quantum fluctuations exist in the SMP. This indicates that the parity-breaking solutions in the SMP can be stable only for a finite time before a large perturbation drives the system away. We thus refer to these parity-breaking solutions as metastable solutions and label the phase as a superradiant metastable phase.

\section{Exact numerical simulation}\label{sec:Exact}
In the preceding section we examined the metastability of the parity-breaking solutions using mean-field theory and the cumulant expansion method. Although these methods provide evidence for the metastable nature of the SMP, they do not clarify its underlying physical origins. To address this limitation, we perform exact numerical simulations of the master equation \eqref{eq:DRM} for finite $\Omega/\omega_0$ and then extrapolate the results to $\Omega/\omega_0\rightarrow \infty$ through finite-size scaling analysis.

\subsection{Steady-state distribution}\label{subSec:SteadystateD}
To further confirm the metastability of the parity-breaking solution in the SMP phase, we resort to exact numerical simulation of the system for finite $\Omega/\omega_0=1200$.  By introducing a truncation of the Fock space of the bosonic mode, we numerically solve the master equation \eqref{eq:DRM}. The steady state is obtained by solving the equation
\begin{equation} \label{eq:steadyState}
\mathcal{L}\rho_s\equiv -i[\hat{H},\rho_s]+\mathcal{D}_{\hat{a}}[\rho_s]+\mathcal{D}_{\hat{\sigma}_{-}}[\rho_s]=0,
\end{equation}
where $\mathcal{L}$ is defined as the Liouvillian of the system. 
The state of the bosonic mode is then obtained by a partial trace of the spin degrees of freedom, defined as $\rho_{B}={\rm{Tr}}_{S}[\rho_s]$. In Fig.~\ref{fig:Qrep} we present the Husimi $Q$ representation of $\rho_B$ at different phases. The $Q$ representation is defined as
\begin{equation}
  Q(\alpha,\alpha^{*})=\frac{1}{\pi} \langle \alpha|\rho_{B}|\alpha\rangle,\label{eq:HusimiQ}
\end{equation}
where $|\alpha\rangle$ is the coherent state of the bosonic mode. It provides a quasiprobability distribution of the bosonic mode in phase space, offering a smooth and non-negative visualization of a quantum state.

In the NP, the $Q$ representation exhibits a single peak centered at the origin, as shown in Fig.~\ref{fig:Qrep}(a), corresponding to the parity-preserving solution predicted by mean-field theory.  The broadening of the peak is attributed to quantum fluctuations.  As the coupling strength $\lambda$ increases beyond the critical point, the $Q$ representation develops two symmetric peaks away from the origin, as shown in Fig.~\ref{fig:Qrep}(b), corresponding to the pair of  parity-breaking solutions predicted by mean-field theory.
The appearance of the paired peaks might suggest that the parity-breaking solutions are stable. However, a closer inspection of the $Q$ representation by rescaling the plot range reveals two distinct channels connecting the parity-breaking solutions and the saddle solution with finite nonvanishing probability, as shown in Fig.~\ref{fig:Qrep}(c). 
Specifically, a weak spiral channel provides a path that can drive the system from the parity-breaking solutions to the saddle point, whereas the sigmoid channel provides a path to settle the system into one of the parity-breaking solutions.
The presence of these channels confirms that the parity-breaking solutions are metastable, consistent with the behavior revealed by the evolution of the symmetry-breaking solution under strong perturbations, as shown in Fig.~\ref{fig:StrongPerturb}(b). Consequently, steady state $\rho_{s}$ in the SMP retains the parity, causing $\langle \hat{x}\rangle_s=\langle \hat{p}\rangle_s=0$ and large quantum fluctuations. It is these enhanced quantum fluctuations that lead to the divergence of higher-order cumulants.

\subsection{Principal component analysis}\label{subSec:PCA}
In the above analysis we assumed that there is a one-to-one correspondence between the steady-state distribution and the dynamics of the parity-breaking solutions. This correspondence arises from the distinctive nature of the steady state in an open quantum system, which is governed by a dynamical equilibrium among multiple contributing components. This stands in contrast to the stationary state of a closed system, which is uniquely determined by the system's Hamiltonian or free energy and typically consists of a single component. When the steady state contains only one single component, it coincides with the stationary state. However, when two or more components coexist, the steady state reflects the dynamical balance between them and thus faithfully reflects the evolution of these components.

\begin{figure}
    \centering
    \includegraphics[width=1.0\linewidth]{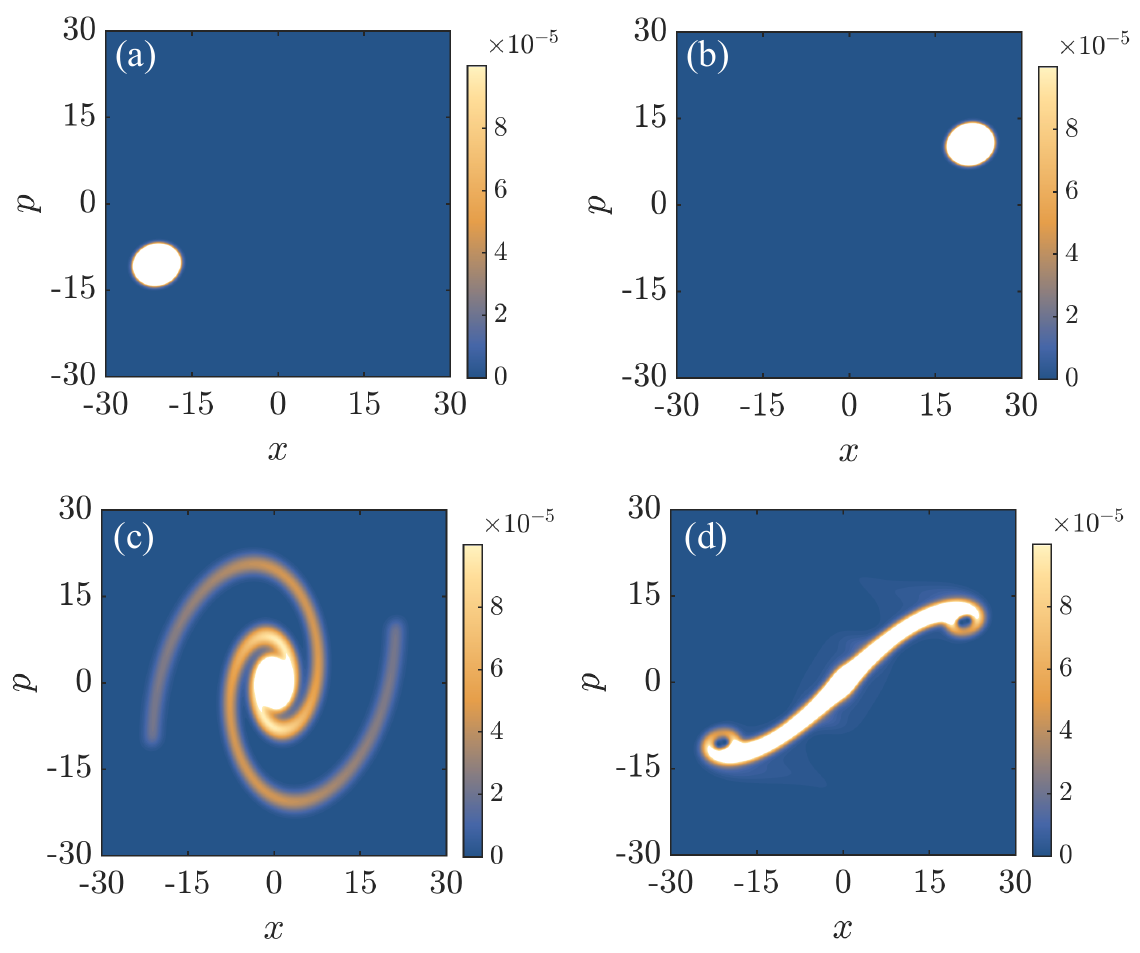}
    \caption{Principal components of the steady state in the SMP. (a) and (b) two parity-breaking components, (c) a spiral and a sigmoid component,  and (d) show a sigmoid component. The parameters are the same as those in Fig.~\ref{fig:cumulant}. }
    \label{fig:PrincipalC}
\end{figure}

To reveal the components in the steady state, we perform an eigenstate decomposition of the steady state,
\begin{equation}\label{eq:ESD}
    \rho_{s}=\sum_{i}p_{i} |\psi_i\rangle \langle \psi_i|,
\end{equation}
where $|\psi_i\rangle$ denotes the $i$th eigenstate with probability $p_i$. Numerical results show that, in the SMP, two states $\vert\psi_{1}\rangle$ and $\vert\psi_{2}\rangle$ dominant with equal probability, corresponding to the parity-breaking solutions, while other states only have quite small probabilities. A further calculation of the $Q$ representation of the bosonic mode reveals that each eigenstate exhibits certain features of the steady-state distribution shown in Fig.~\ref{fig:Qrep}(c) (see Appendix~\ref{sec:Qrep-app} for details). These features can be further classified into different groups according to their similarity.  Such a grouping of the eigenstates yields a principal-component analysis of the steady state
\begin{equation}\label{eq:PCD}
    \rho_s=\sum_{l} \rho^{c}_{l}, \quad \rho^{c}_{l}\equiv \sum_{i \in g_l} p_i |\psi_i\rangle \langle \psi_i|,
\end{equation}
where $\rho^{c}_{l}$ is defined as the principal component $l$ and $g_l$ denotes the set of eigenstates in this group.  Using the standard community detection algorithm developed for graph theory, we find four principal components in $\rho_s$.  They are a pair of parity-breaking components, as shown in Figs.~\ref{fig:PrincipalC}(a) and (b), one spiral component [Fig.~\ref{fig:PrincipalC}(c)], and one sigmoid component [Fig.~\ref{fig:PrincipalC}(d)].  Interpreting the parity-breaking components as metastable states and the spiral and sigmoid components as transient states,  the principal component analysis establishes a correspondence between the steady-state distribution and the dynamics of the metastable states shown in Sec.~\ref{subsec:StablityUP}.  

In the classification, we choose the probability distribution of $\hat{\sigma}_z$ and $\hat{x}$ as the feature of each eigenstate, defined as
\begin{equation}
    P_{\psi_i}(s,x)\equiv |\langle s, x|\psi_i\rangle|^2,
\end{equation}
where $|s,x\rangle$ denotes the simultaneous eigenstate of $\hat{\sigma}_z$ and $\hat{x}$ with eigenvalue $s$ and $x$, respectively. The similarity between two different eigenstates is quantified by the fidelity 
\begin{equation}\label{eq:Fidelity}
    F(P_{\psi_i}, P_{\psi_j})=\left[\sum_{s,x} \sqrt{P_{\psi_i}(s,x)P_{\psi_j}(s,x)}\right]^2.
\end{equation}
Here $0\le F(P_{\psi_i}, P_{\psi_j})\le 1$ with $F=1$ when $P_{\psi_i}=P_{\psi_j}$ and  $F=0$ when $P_{\psi_i}$ and $P_{\psi_j}$ do no overlap.  
We construct a weighted similarity graph of eigenstates based on their similarity, where each eigenstate is represented as a node, and two nodes are connected by an edge weighted by their similarity whenever it exceeds a chosen threshold~\cite{Newman2018}. Applying a standard community detection algorithm to this network with an edge threshold of $0.3$ yields the four principal components shown in Fig.~\ref{fig:PrincipalC}.

\begin{figure}[t]
    \centering
    \includegraphics[width=0.95\linewidth]{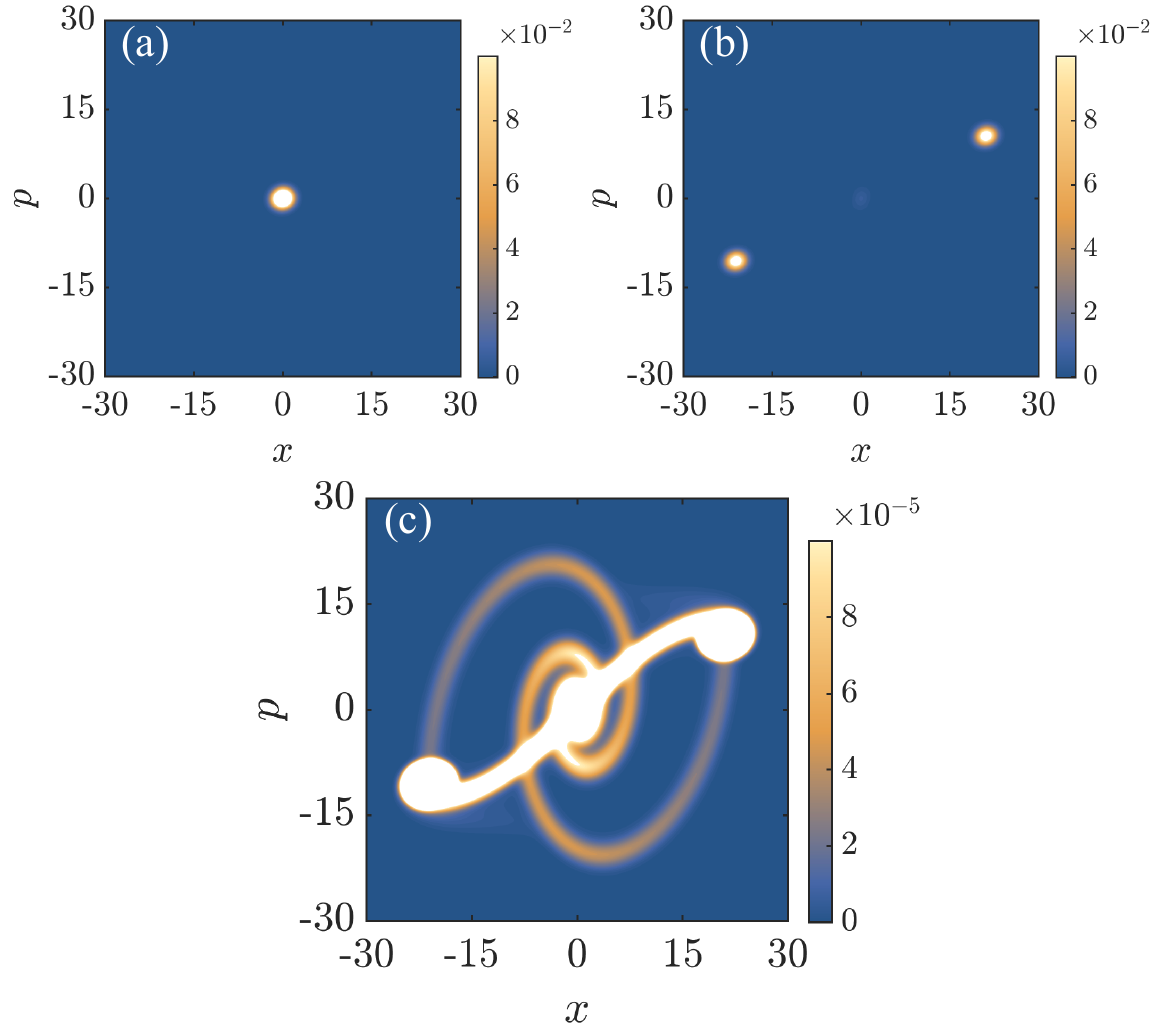}
    \caption{Husimi $Q$ representation of the bosonic mode in the (a) NP and (b) SMP. (c) Magnified view of (b), where the low-intensity regions are emphasized by rescaling the colormap. Here $x\equiv \sqrt{2} \rm{Re}[\alpha]$ and $p\equiv \sqrt{2} \rm{Im}[\alpha]$. The parameters are the same as those in Fig. \ref{fig:cumulant}. }
    \label{fig:Qrep}
\end{figure}

The principal component analysis not only reveals the different components in the steady state but also provides an estimation of the relative lifetime of these components. Owing to the one-to-one correspondence between the steady-state distribution and the dynamic balance between different components, the relative probabilities directly reflect their relative lifetimes, i.e.,
\begin{equation}
    \frac{T_l}{T_{l'}}=\frac{{\rm Tr}[\rho^{c}_{l}]}{{\rm Tr}[\rho^{c}_{l'}]}.
\end{equation}
Numerical results show that the probability of the parity-breaking component is significantly larger than those of the spiral and sigmoid components. Accordingly, the parity-breaking component has a longer lifetime and can thus be regarded as a metastable state, whereas the spiral and sigmoid components are transient. A detailed analysis of the spin probability distribution further reveals that the spin in the spiral component is dominated by the $|+\rangle_{z}$ state, whereas in the sigmoid component it is dominated by the $|-\rangle_{z}$ state [see Fig.~\ref{fig:QTrajectory}(a) for details]. 
Hence, the transition from the spiral component to the sigmoid component is associated with a spin jump, which yields the lifetime $1/2\gamma$ of each spiral channel. Further taking into account the two spiral channels in the spiral component,  the lifetime of the parity-breaking state can then be estimated as
\begin{equation}
    T_{\rm m}=\frac{{\rm Tr}[\rho^{c}_{\rm m}]}{{\rm Tr}[\rho^{c}_{\rm spiral}]} \frac{1}{\gamma},
\end{equation}
where $\rho_{\rm{m}}^{c}$ and $\rho_{\rm{spiral}}^{c}$ denote the parity-breaking  and spiral components, respectively.
For the parameters used in Fig.~\ref{fig:cumulant}, we obtain $\omega_0T_m\approx290$. This characteristic timescale is further confirmed by our quantum trajectory simulation and Liouvillian gap analysis.

\subsection{Quantum trajectory simulation}
\begin{figure*}
    \centering
    \includegraphics[width=0.8\linewidth]{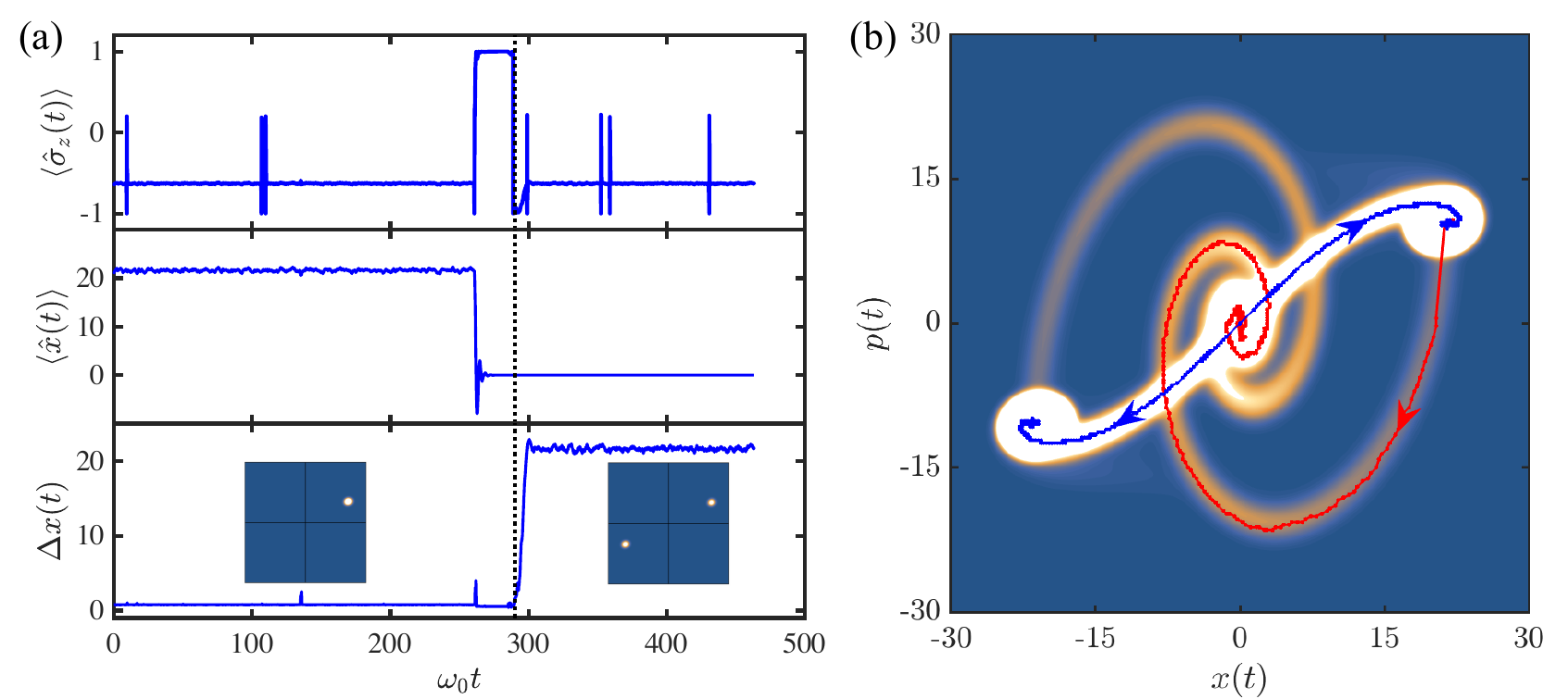}
    \caption{Quantum trajectory of the parity-breaking state $|\psi_1\rangle$ obtained via the quantum jump method. (a) Time evolution of  $\langle \hat{\sigma}_z(t) \rangle$, $\langle \hat{x}(t) \rangle$, and the fluctuation $\Delta x(t) = \sqrt{\langle \hat{x}^2(t) \rangle - \langle \hat{x}(t) \rangle^2}$. The vertical dotted line denotes the time at which the state jumps to the saddle point. Insets display the $Q$ representation of the bosonic mode before and after the saddle point at $\omega_0 t=150.0$ and $\omega_0 t=400.0$, respectively. (b) Evolution path of $\{x (t)$, $p(t)$\} (thick lines) reconstructed by tracking the positions of the peaks in the $Q$ representation of the trajectory state.  The parameters are the same as those in Fig.~\ref{fig:cumulant}. }
    \label{fig:QTrajectory}
\end{figure*}

To gain deeper insight into the correspondence between the dynamics of metastable states and the steady-state distribution, we investigate the system dynamics from an initial parity-breaking state using the quantum jump method~\cite{Dalibard1992, Dum1992, Plenio1998,Link2019}.
In this method, the system state $|\psi(t)\rangle$ evolves unitarily under the Hamiltonian $\hat{H}$, but is interrupted by stochastic quantum jumps induced by the bosonic mode relaxation $\mathcal{D}_{\hat{a}}$ or the spin relaxation $\mathcal{D}_{\hat{\sigma}_-}$. 
The state $|\psi(t)\rangle$ after a small time step $\delta t$ is updated according to
\begin{equation}
    |\psi(t+\delta t)\rangle = \left\{
    \begin{array}{c}
         \sqrt{\frac{2 \kappa \delta t}{P_{\kappa}}}\hat{a} e^{-i \hat{H}\delta t}|\psi(t)\rangle,\\
                \sqrt{\frac{2 \gamma \delta t}{P_{\gamma}}}\hat{\sigma}_{-} e^{-i \hat{H}\delta t}|\psi(t)\rangle, \\
                \frac{1-\gamma \delta t \hat{\sigma}_{+} \hat{\sigma}_{-} -\kappa \delta t \hat{a}^{\dagger} \hat{a}}{1-P_{\kappa}-P_{\gamma}} e^{-i \hat{H}\delta t}|\psi(t)\rangle,\\
    \end{array}\right.
    \label{eq:QJump}
\end{equation}
where the corresponding probabilities for the three possible evolutions are $P_\kappa$, $P_{\gamma}$, and $1-P_{\kappa}-P_{\gamma}$, respectively. Here $P_{\kappa}=2\kappa \delta t \langle \psi(t)|\hat{a}^{\dagger}\hat{a} |\psi(t)\rangle$ and $P_{\gamma}=2\gamma \delta t \langle \psi(t)|\hat{\sigma}_+\hat{\sigma}_- |\psi(t)\rangle$ are the jump probabilities of the bosonic mode and the spin, respectively.
The quantum jump method simulates the dynamics of an open quantum system from the perspective of pure states.
This approach not only provides significant computational advantages but also establishes a direct correspondence with experimental observations in single-system monitoring, such as photon-counting measurements and fluorescence detection in trapped ions. Each stochastic evolution of $|\psi(t)\rangle$ is referred to as a quantum trajectory, and the state of the system $\rho(t)$ is obtained by averaging over all such trajectories.

Taking a representative trajectory as an example, we illustrate the evolution of the parity-breaking state under quantum jumps in the SMP. Starting from a symmetry-breaking state $|\psi_1\rangle$ defined in Eq.~\eqref{eq:ESD}, the quantum trajectory is shown in Fig.~\ref{fig:QTrajectory} by taking $\langle \hat{\sigma}_z(t) \rangle$, $\langle \hat{x}(t) \rangle$, and $\Delta x(t) \equiv \sqrt{\langle \hat{x}^2(t) \rangle-\langle \hat{x}(t) \rangle^2}$ as figures of merit. We find that, in the absence of spin jumps, i.e., sharp spikes in $\langle \hat{\sigma}_z(t) \rangle$, the parity-breaking state remains stable. Each spin jump introduces a strong perturbation to the system, manifesting as a sharp spike in $\langle\hat{\sigma}_z (t)\rangle$. However, in most cases, these perturbations cannot drive the system away from the parity-breaking state, as evidenced by the unaffected evolution of $\langle \hat{x}(t) \rangle$, indicating the robustness of the parity-breaking state. Nevertheless, a small fraction of spin jumps can perturb the system to a transient state with $\langle \hat{\sigma}_z(t)\rangle \approx 1$ and an oscillatory decay of $\langle \hat{x}(t) \rangle$, thereby explicitly demonstrating the metastability of the parity-breaking state. In this case, the system evolves along one of the spiral channel depicted in Fig.~\ref{fig:PrincipalC}(c). A subsequent spin jump (indicated by the vertical dotted line) brings the system to the saddle point characterized by $\langle \hat{\sigma}_z \rangle \approx -1$ and $\langle \hat{x}(t) \rangle \approx 0$. Since the saddle point is unstable, small quantum fluctuations would further cause the system to rapidly relax back to one parity-breaking solution along the sigmoid-shaped channel depicted in Fig.~\ref{fig:PrincipalC}(d). 
Consequently, the trajectory simulation provides direct evidence of the metastable nature of the parity-breaking state.

Notably, the relaxation from the saddle point back to the parity-breaking state is not directly visible in $\langle \hat{x}(t) \rangle$. This is because the parity symmetry is restored at the saddle point, and the subsequent dynamics governed by the quantum jump equation \eqref{eq:QJump} cannot break this symmetry. Consequently, $\langle \hat{\sigma}_x (t)\rangle = 0$ holds throughout the ensuing dynamics.  In this regime, the trajectory state $|\psi(t)\rangle$ forms a cat state, i.e., a superposition of two distinct parity-breaking states, manifested by large fluctuations in $\hat{x}(t)$. This behavior is evident in the evolution of $\Delta x(t)$, which shows a sharp increase after the jump to the saddle point (denoted by the dotted line). The $Q$ representation of the bosonic mode before and after the saddle point, shown in the insets, further illustrates the emergence of the cat state. To obtain the evolution path of the system, we identify the peak in the $Q$ representation and use its position $\{x(t), p(t)\}$ as the representative coordinate.
The reconstructed evolution path $\{x(t),p(t)\}$ is shown as thick lines in Fig.~\ref{fig:QTrajectory}(b). 
A bifurcation of the evolution path (highlighted by blue lines) emerges due to the cat state nature of the $|\psi(t)\rangle$.
The evolution path establishes a one-to-one correspondence between the dynamical evolution of the parity-breaking state and the steady-state distribution, thereby confirming the conclusion in Sec.~\ref{subSec:PCA} and the metastability of the parity-breaking state. 

\section{Lifetime analysis and finite-size scaling}\label{sec:Lifetime}

The principal component analysis provides us a rough estimation of the lifetime of the metastable parity-breaking state. To quantify this lifetime more accurately, we analyze the spectral decomposition of the Liouvillian $\mathcal{L}$ defined in Eq.~\eqref{eq:steadyState}. 
Taking the Liouvillian as a super-operator acting on the density matrix, the spectral decomposition of $\mathcal{L}$ is defined as 
\begin{equation}\label{eq:LiouvillianSpectral}
    \mathcal{L}=\sum_{i=0}\lambda_i |\rho_i \rangle\rangle\langle \langle \tilde{\rho}_i|,
\end{equation}
where $\lambda_i$ is the $i$th eigenvalue and $|\rho_i \rangle\rangle$ and $\langle\langle \tilde{\rho_i}|$ denote the $i$th right and left eigenstates, respectively. Since $\mathcal{L}$ as a matrix is non-Hermitian, the left and right eigenstates are not related by Hermitian conjugation, and the corresponding eigenvalue $\lambda_i$ is in general complex, with the real part representing the decay rate and the imaginary part corresponding to the oscillation frequency~\cite{Kessler2012, Minganti2018}. Here, we assume the eigenvalues are ordered by decreasing real parts such that ${\rm Re}[\lambda_0]\ge {\rm Re}[\lambda_1]\ge {\rm Re}[\lambda_2] \cdots$. One can prove that ${\rm Re}[\lambda_0]=0$ with $|\rho_0\rangle \rangle$ representing the steady state $\rho_s$. From this spectral decomposition, the time evolution of any initial state $\rho(0)$ reads
\begin{equation}
e^{\mathcal{L}t} \rho(0)=\rho_{s}+\sum^{m}_{i=1}c_{i}e^{\lambda_{i}t} \rho_{i} + \sum_{i=m+1}c_{i}e^{\lambda_{i}t} \rho_{i},
\end{equation}
where $\rho_i$ is the matrix representation of $|\rho_i\rangle\rangle$ and the coefficient $c_{i}=\langle \langle \tilde{\rho}_i| \rho(0)\rangle\rangle\equiv {\rm Tr}[\tilde{\rho}_i \rho(0)]$. Here, the sum of $i$ is split into two groups by using the criterion ${\rm Re}[\lambda_m]/{\rm Re}[\lambda_{m+1}] \ll 1$. The subspace spanned by the first $m$ eigenstates constitutes a metastable manifold, characterized by the fact that the eigenstates within it possess long lifetimes~\cite{Macieszczak2016}. 
The dimension of the metastable manifold is related to the symmetry of the Liouvillian. In the dissipative QRM, we the parity symmetry; therefore, only a pair of metastable states are supported, which gives $m=1$. The lifetime of the metastable state is then uniquely determined by the Liouvillian gap $\Delta \equiv {\rm Re}[\lambda_{1}]$ as
\begin{equation}
T_{m}=\Delta^{-1}. 
\end{equation}
For the parameters chosen in Fig.~\ref{fig:Qrep}(c), the numerical calculation yields $\Delta/\omega_0 = 0.0035$, corresponding to $\omega_0 T_m \approx 294.1$. This result is in quantitative agreement with that obtained via principal component analysis and quantum trajectory simulations.

\begin{figure}
    \centering
    \includegraphics[width=1.0 \linewidth]{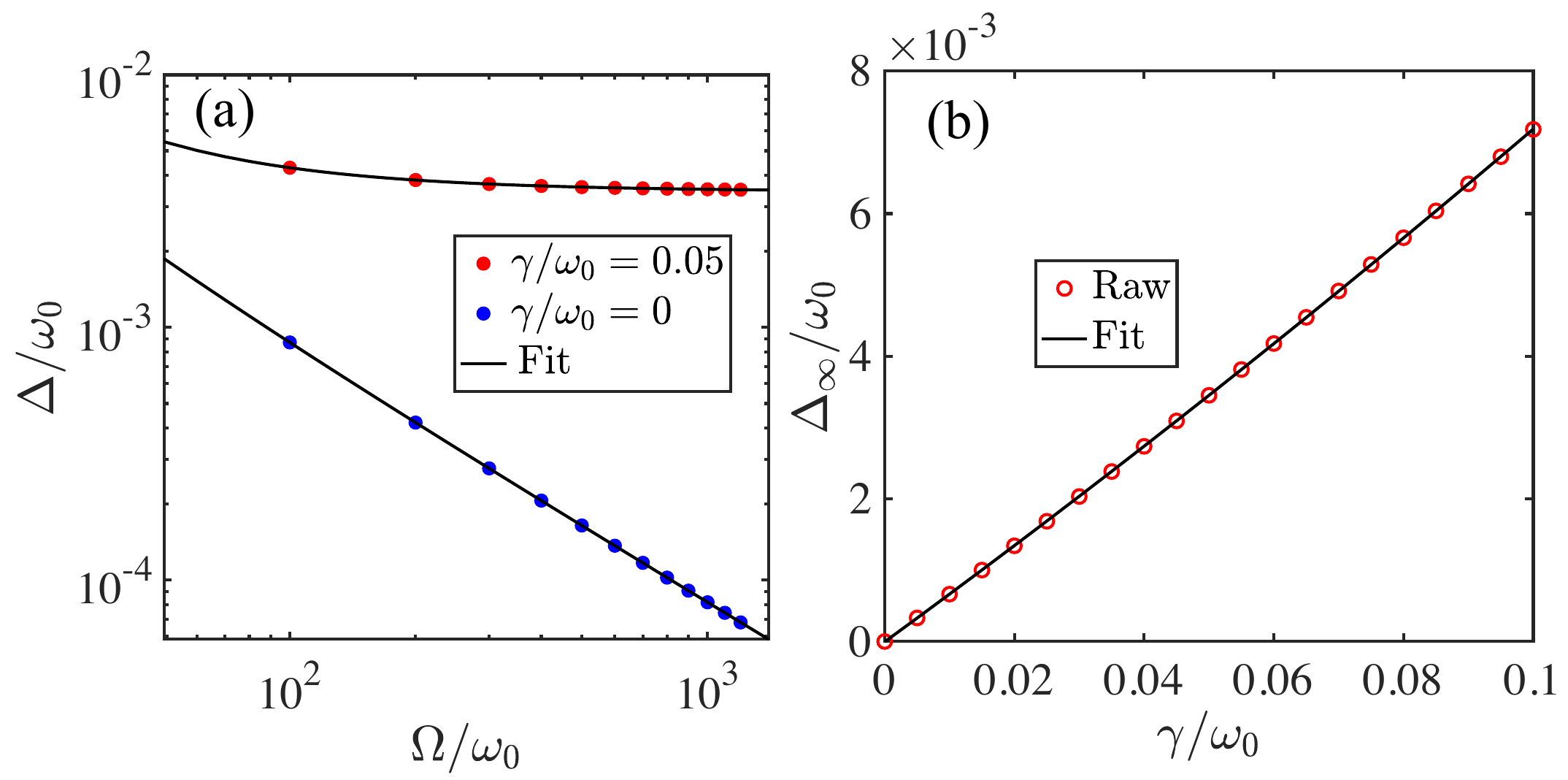}
    \caption{(a) Finite-size scaling of the Liouvillian gap with $\lambda=1.4\sqrt{{\Omega \omega_0}/{2}}$ and (b) dependence of the Liouvillian gap $\Delta_{\infty}$ on the spin relaxation rate $\gamma$ in the thermodynamic limit. For (a), the numerical fitting $a+b \frac{\omega_0}{\Omega}+c\frac{\omega^2_0}{\Omega^2}$ yields coefficients $a=0.0034, b=0.068, c=1.54$ for $\gamma/\omega_0=0.05$ and $a=0.0,b=0.081,c=0.64$ for $\gamma/\omega_0=0$. For (b), the numerical fitting $a+b\frac{\gamma}{\omega_0}+c\frac{\gamma^2}{\omega_0^2}$ yields coefficients $a=0.0,b=0.066,c=0.055$.
    The other parameters are the same as those in Fig.~\ref{fig:cumulant}.}
    \label{fig:scaling}
\end{figure}

The above numerical results are obtained for finite $\Omega/\omega_0$. It is necessary to examine the metastability of the parity-breaking solutions in the thermodynamic limit, defined by $\Omega/\omega_0 \to \infty$~\cite{Emary2003,Emary2003A}. In particular, we need to examine whether the Liouvillian gap closes in this limit. According to Liouvillian spectral theory, the gap closes in the symmetry-breaking phase in the thermodynamic limit~\cite{Minganti2018}, since a symmetry-breaking steady state implies that all its symmetry-transformed counterparts are also steady states. Thus, the Liouvillian contains at least two eigenstates with zero eigenvalues. Conversely, if the Liouvillian gap remains open in the thermodynamic limit, the symmetry-breaking phase can only be metastable. A criterion of the SMP is then obtained as
\begin{equation}\label{eq:criteria}
    \Delta_\infty =\left\{ 
    \begin{array}{cc}
        0, & \text{SP},  \\
        \text{finite}, & \text{SMP}, 
    \end{array}
    \right.
\end{equation}
where $\Delta_\infty\equiv \lim_{\Omega/\omega_0
\rightarrow \infty} \Delta$ denotes the gap in the thermodynamic limit.

To assess the Liouvillian gap $\Delta_\infty$, we perform finite-size scaling with fixed $\lambda^2/\Omega \omega_0$.
The results, presented in Fig.~\ref{fig:scaling}(a), show that the Liouvillian gap $\Delta$ decreases very slowly with increasing $\Omega/\omega_0$. A polynomial fit $\Delta=a+b \frac{\omega_0}{\Omega}+c\frac{\omega^2_0}{\Omega^2}$ yields that the gap approaches a finite value $\Delta_{\infty}\approx 0.0034$ as $\Omega/\omega_0 \to \infty$, confirming the metastability of the parity-breaking states in the thermodynamic limit. For comparison, we also analyze the scaling of the gap at $\gamma=0$, which exhibits a power-law closing as $\Omega/\omega_0$ increases. This behavior indicates that the SP becomes stable in the absence of spin relaxation. However, the power-law closing, in contrast to the exponential closing of typical equilibrium phase transitions, suggests that the SP at $\gamma=0$ should still be regarded as metastable in a broader sense. A more precise characterization of the distinction between power-law and exponential gap closing is beyond the scope of this work.
In Fig.~\ref{fig:scaling}(b) we show the dependence of the Liouvillian gap $\Delta_{\infty}$ on the spin relaxation rate $\gamma$, which reveals that $\Delta_{\infty}$ vanishes linearly with decreasing $\gamma$. This indicates that the SP is stable (metastable in a broader sense) only in the $\gamma=0$ limit. Even a small spin relaxation would render the SP a SMP, which explains the phase diagram in Fig.~\ref{fig:Scheme}(b).

\section{Discussion}\label{sec:Discussion}
The above analysis provides a comprehensive investigation of the metastability of the parity-breaking states in the SMP of the dissipative QRM. In this model, spin relaxation induces random spin flips at a fixed rate. Each flip acts as a sudden quench to the system, introducing a strong perturbation that drives the system out of the linear stability regime of the parity-breaking state into the parity-preserving state with a finite probability. Owing to the saddle-point nature of the parity-preserving state, the system returns to one of the parity-breaking states. This recycling process underlies the metastability of the SMP and restores the parity symmetry of the steady state.

Metastability is a common feature in dissipative phase transitions~\cite{Haken1986,Letscher2017, Rota2018,Gerry2024}. For instance, in the lasing phase transition, the system transits from the normal phase to the lasing phase characterized by spontaneous breaking of the global U(1) symmetry above the lasing threshold. However, atomic pumping, which plays a role analogous to spin relaxation, continuously quenches the system and introduces perturbations that prevent the laser maintaining a definite phase. This results in phase diffusion that gradually restores the global U(1) symmetry, rendering the lasing phase metastable. 

The stability analysis developed for the dissipative QRM can be generalized to other open quantum systems with dissipative phase transitions. In this paper we showed that the linear stability analysis only characterizes the stability of the symmetry-breaking state under small perturbations. It is essential to assess whether quantum fluctuations are negligible, for example, by examining the convergence of the cumulant expansion. Another approach to evaluate the stability of the symmetry-breaking phase is to investigate the closing of the Liouvillian gap in the thermodynamic limit. However, for most many-body systems, calculating the Liouvillian gap is challenging, as in the case of lasing models. For such systems, the semiclassical Fokker-Planck equation can be used to estimate the phase diffusion rate, which is directly related to the Liouvillian gap.

Through the study of the dissipative QRM, we can draw a comparison between phase transitions in equilibrium and open quantum systems. In equilibrium systems, phase transitions occur when the equilibrium state changes from a symmetry-preserving state to a symmetry-breaking state. This symmetry-breaking state typically corresponds to the global minimum of the free energy or the ground state energy. Although Goldstone modes exist as zero-energy excitations when a continuous symmetry is broken, the Mermin-Wagner theorem guarantees that such excitations can be neglected in high-dimensional systems~\cite{Mermin1966}. Therefore, the symmetry-breaking state in equilibrium systems is generally stable. In contrast, dissipative phase transition in open quantum systems exhibit fundamentally different behaviors. First, the steady state corresponds to local minima of an effective potential, which can give rise to multistable phases. Second, relaxation or driven processes, unavoidable in open systems, may act as random and strong quenches to the system. These processes can drive the system from one local minimum to another, making the resulting steady state determined by a dynamical balance between different components in the steady state and leading to the metastability.

Finally, a variety of experimental platforms can be used to verify the SMP of the dissipative QRM under a weak spin relaxation, such as trapped ions~\cite{Puebla2017,Lv2018,Cai2021}, nuclear magnetic resonance ~\cite{Chen2021, Wu2024}, and superconducting circuits~\cite{Braumueller2017, Langford2017}. Notably, the dissipative phase transition at $\gamma=0$ has been demonstrated in recent experiments~\cite{Wu2024}. We expect that the metastability of the superradiant phase can be experimentally tested on these platforms.

\section{Conclusion}\label{sec:Conclusion}

In this paper we investigated the metastability of the dissipative phase transition of the QRM in the presence of a weak spin relaxation. We revealed that the symmetry-breaking states in this model are inherently metastable. Spin relaxation induces random spin flips, which further act as strong perturbations to the system, facilitating transitions between different symmetry-breaking states. These transitions ultimately restore the system symmetry and give rise to the metastability of the symmetry-breaking phase.

To elucidate the metastability of the dissipative QRM, we performed both mean-field and cumulant expansion analyses, as well as exact numerical simulations for finite $\Omega/\omega_0$. These revealed a one-to-one correspondence between the steady-state distribution and the dynamic evolution of the metastable state. 
Based on this correspondence, we identified the coexistence of different components in the SMP and provide a rough estimate of their relative lifetimes from the principal component analysis.
To quantify the lifetime more precisely and explore the fate of the superradiant metastable phase in the thermodynamic limit, we analyzed the Liouvillian spectrum. The results show that the lifetime is uniquely governed by the Liouvillian gap and is also consistent with the principal component analysis. Finite-size scaling further confirmed that the superradiant phase is stable only in the absence of spin relaxation but becomes metastable once a small spin relaxation is present.

Metastable symmetry-breaking phases are a common feature of open quantum systems and underscore a fundamental distinction between phase transitions in equilibrium and nonequilibrium systems. Our findings offer insights into the metastability of dissipative phase transitions in open quantum systems.

\section{Data Availability} 
The data that support the findings of this article are not publicly available. The data are available from the authors upon reasonable request.

\appendix

\section{Cumulant expansion method}\label{sec:CEM-app}

From the master equation \eqref{eq:DRM}, we can derive the equation of motion for any operator $\hat{O}=\hat{o}^{n_1}_i \hat{o}^{n_2}_j \cdots$ (i.e., correlations) as 
\begin{equation}\label{eq:OperatorEq}
   \frac{d}{dt}\langle \hat{O} \rangle= i \langle [\hat{H}, \hat{O}] \rangle +  \langle \mathcal{D}'_{\hat{a}}[\hat{O}] \rangle+\langle \mathcal{D}'_{\hat{\sigma}_{+}}[\hat{O}] \rangle,
\end{equation}
with $\mathcal{D}'_{\hat{a}}[\hat{O}]\equiv 2\kappa \hat{a}^{\dagger} \hat{O}\hat{a}- \kappa  \hat{a}^{\dagger} \hat{a}\hat{O}-\kappa  \hat{O}\hat{a}^{\dagger} \hat{a}$ and  $\mathcal{D}'_{\hat{\sigma}_{-}}[\hat{O}]\equiv 2\gamma \hat{\sigma}_{+} \hat{O}\hat{\sigma}_{-}- \gamma \hat{\sigma}_{+}\hat{\sigma}_{-}\hat{O}-\gamma \hat{O} \hat{\sigma}_{+}\hat{\sigma}_{-}$. 
The order of the correlations is defined as $n \equiv n_1+n_2+\cdots$, and the operators $\hat{o}_i$ can be $\hat{x}$, $\hat{p}$, or $\hat{\sigma}_{x,y,z}$ in the dissipative QRM.
A specific example is 
\begin{equation}
\begin{aligned}
    \frac{d}{dt}\langle \hat{\sigma}_y \hat{p} \rangle= &-(\kappa+\gamma) \langle \hat{\sigma}_y \hat{p} \rangle-\omega_0\langle \hat{\sigma}_y \hat{x} \rangle+\Omega \langle \hat{\sigma}_x \hat{p}  \rangle \\
    &-\lambda  \langle \hat{\sigma}_z (\hat{x}\hat{p}+\hat{p}\hat{x})\rangle.
\end{aligned}
\end{equation}
It can be observed that the equations for second-order correlations couple to third-order correlations, giving rise to a hierarchical structure in which low-order correlations are coupled to higher-order ones. This hierarchy leads to an infinite set of coupled equations, thereby precluding an exact solution of the correlations.

Here we employ the cumulant expansion method to truncate the hierarchy equations and close these correlation equations~\cite{Barquilla2020, Leymann2014,plankensteiner2022}. The cumulant $\langle \hat{o}^{n_1}_{i_1} \hat{o}^{n_2}_{i_2} \cdots \rangle_c$ is defined from the generating function $\langle e^{\sum_{l} \eta_l \hat{o}_l}\rangle$ as
\begin{equation}
    \langle \hat{o}^{n_1}_{i_1} \hat{o}^{n_2}_{i_2} \cdots \rangle_c\equiv \frac{d^{n_1}}{d \eta^{n_1}_{i_1}} \frac{d^{n_{2}}}{d\eta^{n_2}_{i_2}} \cdots \left. \ln \langle e^{\sum_{l} \eta_l \hat{o}_l}\rangle\right|_{\vec{\eta}=\vec{0}},
\end{equation}
where $n=n_1+n_2+\cdots$ denotes the order of the cumulant~\cite{Kubo1962}. Through the definition of cumulants, one can establish relations between correlations and cumulants. For low-order correlations, we have
\begin{align}
    \langle \hat{o}_i\rangle=&\langle o_i \rangle_c, \\
     \langle \hat{o}_i \hat{o}_j\rangle=&\langle \hat{o}_i \hat{o}_j\rangle_c+\langle \hat{o}_i  \rangle_c\langle  \hat{o}_j \rangle_c, \\
    \langle \hat{o}_i \hat{o}_j \hat{o}_k\rangle=&\langle\hat{o}_i \hat{o}_j \hat{o}_k \rangle_c+ \langle \hat{o}_i \hat{o}_j\rangle_c \langle \hat{o}_k \rangle_c +\langle \hat{o}_i \hat{o}_k\rangle_c \langle \hat{o}_j \rangle_c, \nonumber \\
    &+\langle\hat{o}_i\rangle_c \langle \hat{o}_j \hat{o}_k\rangle_c+
    \langle\hat{o}_i\rangle_c \langle\hat{o}_j\rangle_c \langle\hat{o}_k \rangle_c.
\end{align}
One can observe that the $n$th order correlations depend only on cumulants up to $n$th order.

Applying the cumulant expansion to both sides of Eq.~\eqref{eq:OperatorEq}, we obtain the equations of motion for cumulants. The left-hand side includes correlations (and hence cumulants) up to $n$th order, while the right-hand side includes correlations up to $(n+1)$th order due to the hierarchy structure. By truncating the hierarchical through neglecting these $(n+1)$th order cumulants on the right-hand side, the equations of motion for all operators $\{\hat{O}(t)\}$ up to $n$th order become closed in terms of cumulants, enabling us to solve them.

A major advantage of the cumulant expansion method over naive truncation of correlations lies in its systematic treatment of genuine coupled correlations~\cite{Kubo1962}.  Cumulants quantify the intrinsic correlations between operators, which generally decrease as the order increases. By neglecting the $(n+1)$th order cumulants, all correlations up to the $n$th order are retained exactly, while all higher-order correlations are approximated in a self-consistent manner. For example, the second-order truncation yields $\langle \hat{o}_i \hat{o}_j \hat{o}_k\rangle \approx \langle \hat{o}_i \hat{o}_j \rangle_c \langle \hat{o}_k \rangle_c + \langle \hat{o}_i \hat{o}_k\rangle_c \langle \hat{o}_j \rangle_c + \langle\hat{o}_i\rangle_c \langle \hat{o}_j \hat{o}_k\rangle_c + 
    \langle\hat{o}_i\rangle_c $ $ \langle\hat{o}_j\rangle_c \langle\hat{o}_k \rangle_c$. This truncation provides a good approximation to $\langle \hat{o}_i \hat{o}_j \hat{o}_k\rangle$ when $\langle \hat{o}_i \hat{o}_j \hat{o}_k\rangle_c$ is negligible.
Thus, the cumulant expansion method provides a controlled and physically motivated approximation that improves both the accuracy and stability of the resulting equations of motion.

\section{Spectral decomposition of the steady state}\label{sec:Qrep-app}
\begin{figure}
    \centering
    \includegraphics[width=0.9\linewidth]{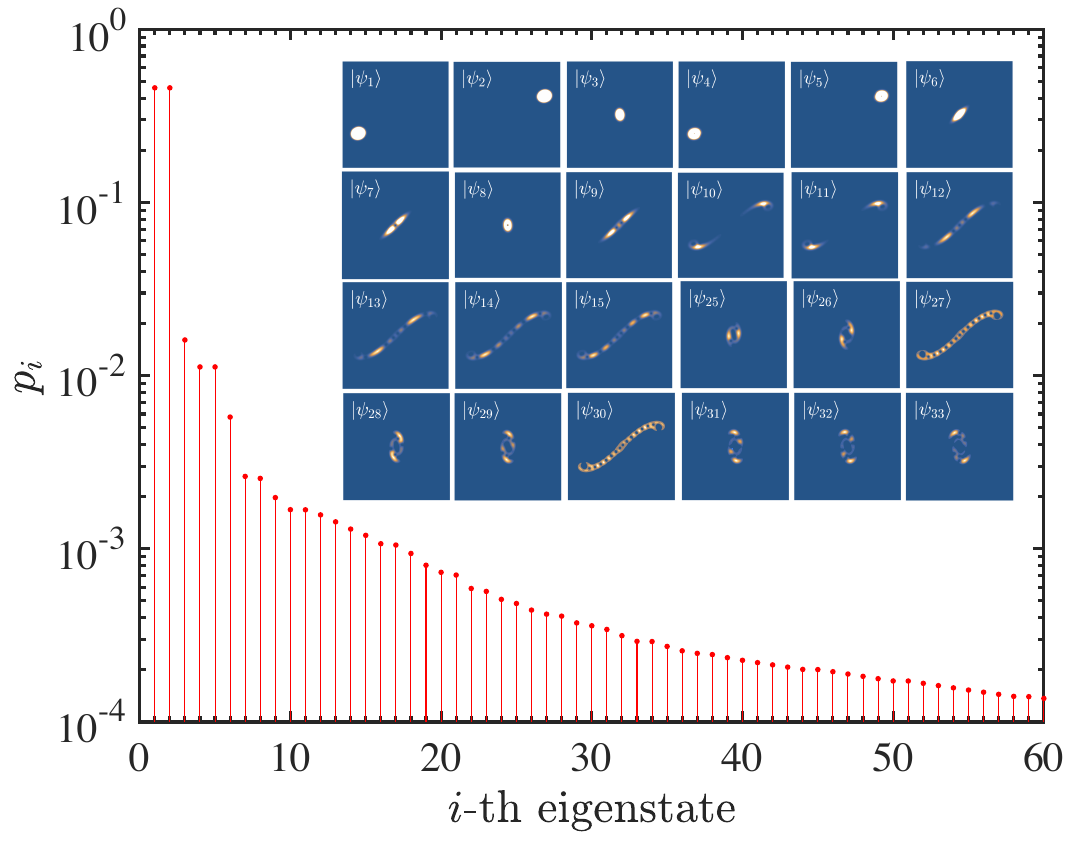}
    \caption{ Distribution of $p_i$ in the spectral decomposition of the steady state. The inset shows the $Q$ representations of the bosonic mode in the eigenstates $|\psi_i\rangle$ of the steady state.
    }
    \label{fig:PEig-app}
\end{figure}

In the main text, we introduced an eigenstate decomposition of the steady state,
\begin{equation}
    \rho_s=\sum_{i} p_i |\psi_{i}\rangle \langle \psi_i|,
\end{equation}
where $|\psi_i\rangle$ denotes the $i$th eigenstate with probability $p_i$. Assuming that the probabilities ${p_i}$ are ordered in descending order, their distribution is shown in Fig.~\ref{fig:PEig-app}. Two dominant eigenstates $|\psi_1\rangle$ and $|\psi_2\rangle$ have equal probabilities. Their $Q$ representations, displayed in the inset, confirm that they correspond to the pair of parity-breaking states. For each eigenstate, the $Q$ representation of the bosonic mode is used as its characteristic signature.  Representative examples are shown in the inset, revealing that each eigenstate captures certain features of the steady-state distribution. Moreover, several eigenstates share similar features, suggesting that they can be grouped into the same classes. As shown in the main text, we group them into four classes using a standard community detection algorithm in graph theory.

\section{steady state solution}\label{sec:conv-check}
The steady state of Eq.~(\ref{eq:steadyState}) is obtained by numerically  solving the linear equation
\begin{equation}
   \sum_{\alpha,\alpha'} L_{\tilde{\alpha},\tilde{\alpha}'; \alpha,\alpha'} R_{\alpha,\alpha'}=0,
\end{equation}
where $L$ and $R$ denote the Liouvillian and the steady state in the Liouville-space basis
\begin{align}
    \rho_{s}=&\sum_{\alpha,\alpha'} R_{\alpha, \alpha'}\vert \alpha,\alpha'\rangle\rangle,\\
    \mathcal{L}=&\sum_{\alpha,\alpha', \tilde{\alpha},\tilde{\alpha}'} L_{\tilde{\alpha},\tilde{\alpha}'; \alpha,\alpha'} \vert  \tilde{\alpha},\tilde{\alpha}'\rangle\rangle  \langle\langle\alpha, \alpha' \vert.
\end{align}
Here, $\alpha=\{s,n\}$, where $s$ and $n$ label the spin up or down and the photon number, respectively, $R_{\alpha,\alpha'}\equiv \langle \alpha' |\rho|\alpha\rangle$, and $L_{\tilde{\alpha},\tilde{\alpha}'; \alpha,\alpha'} \equiv \rm{Tr}[(|\tilde{\alpha}'\rangle \langle \tilde{\alpha}|) \mathcal{L} (|\alpha\rangle \langle \alpha'|)]$. In the numerical calculation, we impose a photon number cutoff $n\le N_{\rm cut}$. Convergence is verified by varying $N_{\rm cut}$, which shows that $\rho_s$ saturates for $N_{\rm cut}=499$ with a residual error below $10^{-12}$.

\bibliography{DissipativeQRM}

@article{Kubo1962,
author = {Kubo ,Ryogo},
title = {Generalized Cumulant Expansion Method},
journal = {J. Phys. Soc. Jpn.},
volume = {17},
number = {7},
pages = {1100-1120},
year = {1962},
doi = {10.1143/JPSJ.17.1100},

URL = {https://doi.org/10.1143/JPSJ.17.1100}
}

@article{plankensteiner2022,
  title={Quantum{C}umulants. jl: A Julia framework for generalized mean-field equations in open quantum systems},
  author={Plankensteiner, David and Hotter, Christoph and Ritsch, Helmut},
  journal={Quantum},
  volume={6},
  pages={617},
  year={2022},
  publisher={Verein zur F{\"o}rderung des Open Access Publizierens in den Quantenwissenschaften},
  doi={10.22331/q-2022-01-04-617},
  URL={https://doi.org/10.22331/q-2022-01-04-617}
}

@article{Dalibard1992,
  title = {Wave-function approach to dissipative processes in quantum optics},
  author = {Dalibard, Jean and Castin, Yvan and M\o{}lmer, Klaus},
  journal = {Phys. Rev. Lett.},
  volume = {68},
  issue = {5},
  pages = {580--583},
  numpages = {0},
  year = {1992},
  month = {Feb},
  publisher = {American Physical Society},
  doi = {10.1103/PhysRevLett.68.580},
  url = {https://link.aps.org/doi/10.1103/PhysRevLett.68.580}
}

@article{Dum1992,
  title = {Monte Carlo simulation of the atomic master equation for spontaneous emission},
  author = {Dum, R. and Zoller, P. and Ritsch, H.},
  journal = {Phys. Rev. A},
  volume = {45},
  issue = {7},
  pages = {4879--4887},
  numpages = {0},
  year = {1992},
  month = {Apr},
  publisher = {American Physical Society},
  doi = {10.1103/PhysRevA.45.4879},
  url = {https://link.aps.org/doi/10.1103/PhysRevA.45.4879}
}

@article{Plenio1998,
  title = {The quantum-jump approach to dissipative dynamics in quantum optics},
  author = {Plenio, M. B. and Knight, P. L.},
  journal = {Rev. Mod. Phys.},
  volume = {70},
  issue = {1},
  pages = {101--144},
  numpages = {0},
  year = {1998},
  month = {Jan},
  publisher = {American Physical Society},
  doi = {10.1103/RevModPhys.70.101},
  url = {https://link.aps.org/doi/10.1103/RevModPhys.70.101}
}

@article{Braak2011,
  title = {Integrability of the Rabi Model},
  author = {Braak, D.},
  journal = {Phys. Rev. Lett.},
  volume = {107},
  issue = {10},
  pages = {100401},
  numpages = {4},
  year = {2011},
  month = {Aug},
  publisher = {American Physical Society},
  doi = {10.1103/PhysRevLett.107.100401},
  url = {https://link.aps.org/doi/10.1103/PhysRevLett.107.100401}
}

@article{Kessler2012,
  title = {Dissipative phase transition in a central spin system},
  author = {Kessler, E. M. and Giedke, G. and Imamoglu, A. and Yelin, S. F. and Lukin, M. D. and Cirac, J. I.},
  journal = {Phys. Rev. A},
  volume = {86},
  issue = {1},
  pages = {012116},
  numpages = {21},
  year = {2012},
  month = {Jul},
  publisher = {American Physical Society},
  doi = {10.1103/PhysRevA.86.012116},
  url = {https://link.aps.org/doi/10.1103/PhysRevA.86.012116}
}

@article{Barquilla2020,
    author = {Sánchez-Barquilla, M. and Silva, R. E. F. and Feist, J.},
    title = {Cumulant expansion for the treatment of light–matter interactions in arbitrary material structures},
    journal = {J. Chem. Phys.},
    volume = {152},
    number = {3},
    pages = {034108},
    year = {2020},
    month = {01},
    issn = {0021-9606},
    doi = {10.1063/1.5138937},
    url = {https://doi.org/10.1063/1.5138937}
}

@article{Leymann2014,
  title = {Expectation value based equation-of-motion approach for open quantum systems: A general formalism},
  author = {Leymann, H. A. M. and Foerster, A. and Wiersig, J.},
  journal = {Phys. Rev. B},
  volume = {89},
  issue = {8},
  pages = {085308},
  numpages = {11},
  year = {2014},
  month = {Feb},
  publisher = {American Physical Society},
  doi = {10.1103/PhysRevB.89.085308},
  url = {https://link.aps.org/doi/10.1103/PhysRevB.89.085308}
}

@article{Hwang2015,
  title = {Quantum {{Phase Transition}} and {{Universal Dynamics}} in the {{Rabi Model}}},
  author = {Hwang, Myung-Joong and Puebla, Ricardo and Plenio, Martin B.},
  year = {2015},
  month = oct,
  journal = {Phys. Rev. Lett.},
  volume = {115},
  number = {18},
  pages = {180404},
  issn = {0031-9007, 1079-7114},
  doi = {10.1103/PhysRevLett.115.180404},
  urldate = {2025-08-02},
  copyright = {http://link.aps.org/licenses/aps-default-license},
  langid = {english}
}

@article{Liu2017,
  title = {Universal Scaling and Critical Exponents of the Anisotropic Quantum Rabi Model},
  author = {Liu, Maoxin and Chesi, Stefano and Ying, Zu-Jian and Chen, Xiaosong and Luo, Hong-Gang and Lin, Hai-Qing},
  journal = {Phys. Rev. Lett.},
  volume = {119},
  issue = {22},
  pages = {220601},
  numpages = {6},
  year = {2017},
  month = {Nov},
  publisher = {American Physical Society},
  doi = {10.1103/PhysRevLett.119.220601},
  url = {https://link.aps.org/doi/10.1103/PhysRevLett.119.220601}
}

@article{Hwang2018,
  title = {Dissipative Phase Transition in the Open Quantum {{Rabi}} Model},
  author = {Hwang, Myung-Joong and Rabl, Peter and Plenio, Martin B.},
  year = {2018},
  month = jan,
  journal = {Phys. Rev. A},
  volume = {97},
  number = {1},
  pages = {013825},
  issn = {2469-9926, 2469-9934},
  doi = {10.1103/PhysRevA.97.013825},
  urldate = {2025-08-02},
  langid = {english},
}

@article{Bakemeier2012,
  title = {Quantum phase transition in the Dicke model with critical and noncritical entanglement},
  author = {Bakemeier, L. and Alvermann, A. and Fehske, H.},
  journal = {Phys. Rev. A},
  volume = {85},
  issue = {4},
  pages = {043821},
  numpages = {9},
  year = {2012},
  month = {Apr},
  publisher = {American Physical Society},
  doi = {10.1103/PhysRevA.85.043821},
  url = {https://link.aps.org/doi/10.1103/PhysRevA.85.043821}
}

@article{Minganti2018,
  title = {Spectral theory of Liouvillians for dissipative phase transitions},
  author = {Minganti, Fabrizio and Biella, Alberto and Bartolo, Nicola and Ciuti, Cristiano},
  journal = {Phys. Rev. A},
  volume = {98},
  issue = {4},
  pages = {042118},
  numpages = {13},
  year = {2018},
  month = {Oct},
  publisher = {American Physical Society},
  doi = {10.1103/PhysRevA.98.042118},
  url = {https://link.aps.org/doi/10.1103/PhysRevA.98.042118}
}

@article{Macieszczak2016,
  title = {Towards a Theory of Metastability in Open Quantum Dynamics},
  author = {Macieszczak, Katarzyna and Guta, Madalin and Lesanovsky, Igor and Garrahan, Juan P.},
  journal = {Phys. Rev. Lett.},
  volume = {116},
  issue = {24},
  pages = {240404},
  numpages = {6},
  year = {2016},
  month = {Jun},
  publisher = {American Physical Society},
  doi = {10.1103/PhysRevLett.116.240404},
  url = {https://link.aps.org/doi/10.1103/PhysRevLett.116.240404}
}

@article{Rose2016,
  title = {Metastability in an open quantum Ising model},
  author = {Rose, Dominic C. and Macieszczak, Katarzyna and Lesanovsky, Igor and Garrahan, Juan P.},
  journal = {Phys. Rev. E},
  volume = {94},
  issue = {5},
  pages = {052132},
  numpages = {11},
  year = {2016},
  month = {Nov},
  publisher = {American Physical Society},
  doi = {10.1103/PhysRevE.94.052132},
  url = {https://link.aps.org/doi/10.1103/PhysRevE.94.052132}
}

@article{Macieszczak2021,
  title = {Theory of classical metastability in open quantum systems},
  author = {Macieszczak, Katarzyna and Rose, Dominic C. and Lesanovsky, Igor and Garrahan, Juan P.},
  journal = {Phys. Rev. Res.},
  volume = {3},
  issue = {3},
  pages = {033047},
  numpages = {26},
  year = {2021},
  month = {Jul},
  publisher = {American Physical Society},
  doi = {10.1103/PhysRevResearch.3.033047},
  url = {https://link.aps.org/doi/10.1103/PhysRevResearch.3.033047}
}

@article{Rota2018,
doi = {10.1088/1367-2630/aab703},
url = {https://dx.doi.org/10.1088/1367-2630/aab703},
year = {2018},
month = {apr},
publisher = {IOP Publishing},
volume = {20},
number = {4},
pages = {045003},
author = {Rota, R and Minganti, F and Biella, A and Ciuti, C},
title = {Dynamical properties of dissipative XYZ Heisenberg lattices},
journal = {New J. Phys.}
}

@article{Mermin1966,
  title = {Absence of Ferromagnetism or Antiferromagnetism in One- or Two-Dimensional Isotropic Heisenberg Models},
  author = {Mermin, N. D. and Wagner, H.},
  journal = {Phys. Rev. Lett.},
  volume = {17},
  issue = {22},
  pages = {1133--1136},
  numpages = {0},
  year = {1966},
  month = {Nov},
  publisher = {American Physical Society},
  doi = {10.1103/PhysRevLett.17.1133},
  url = {https://link.aps.org/doi/10.1103/PhysRevLett.17.1133}
}

@article{Emary2003,
  title = {Chaos and the quantum phase transition in the Dicke model},
  author = {Emary, Clive and Brandes, Tobias},
  journal = {Phys. Rev. E},
  volume = {67},
  issue = {6},
  pages = {066203},
  numpages = {22},
  year = {2003},
  month = {Jun},
  publisher = {American Physical Society},
  doi = {10.1103/PhysRevE.67.066203},
  url = {https://link.aps.org/doi/10.1103/PhysRevE.67.066203}
}

@article{Emary2003A,
  title = {Quantum Chaos Triggered by Precursors of a Quantum Phase Transition: The Dicke Model},
  author = {Emary, Clive and Brandes, Tobias},
  journal = {Phys. Rev. Lett.},
  volume = {90},
  issue = {4},
  pages = {044101},
  numpages = {4},
  year = {2003},
  month = {Jan},
  publisher = {American Physical Society},
  doi = {10.1103/PhysRevLett.90.044101},
  url = {https://link.aps.org/doi/10.1103/PhysRevLett.90.044101}
}

@book{Newman2018,
    author = {Newman, Mark},
    title = {Networks},
    publisher = {Oxford University Press},
    year = {2018},
    month = {07},
    isbn = {9780198805090},
    doi = {10.1093/oso/9780198805090.001.0001},
    url = {https://doi.org/10.1093/oso/9780198805090.001.0001},
}

@book{Haken1986,
  title={Laser {L}ight {D}ynamics},
  author={Haken, Hermann},
  year={1986},
  publisher={North-Holland Amsterdam}
}

@Article{Link2019,
  author       = {Link, Valentin and Luoma, Kimmo and Strunz, Walter T.},
  journal      = {Phys. Rev. A},
  title        = {Revealing the nature of nonequilibrium phase transitions with quantum trajectories},
  year         = {2019},
  month        = {Jun},
  pages        = {062120},
  volume       = {99},
  creationdate = {2025-08-18T01:57:32},
  doi          = {10.1103/PhysRevA.99.062120},
  issue        = {6},
  numpages     = {7},
  publisher    = {American Physical Society},
  url          = {https://link.aps.org/doi/10.1103/PhysRevA.99.062120},
}

@Article{Rabi1936,
  author       = {Rabi, I. I.},
  journal      = {Phys. Rev.},
  title        = {On the Process of Space Quantization},
  year         = {1936},
  month        = {Feb},
  pages        = {324--328},
  volume       = {49},
  creationdate = {2025-08-19T15:12:05},
  doi          = {10.1103/PhysRev.49.324},
  issue        = {4},
  numpages     = {0},
  publisher    = {American Physical Society},
  url          = {https://link.aps.org/doi/10.1103/PhysRev.49.324},
}

@Article{Yoshihara2016,
  author       = {Yoshihara, Fumiki and Fuse, Tomoko and Ashhab, Sahel and Kakuyanagi, Kosuke and Saito, Shiro and Semba, Kouichi},
  journal      = {Nat. Phys.},
  title        = {Superconducting qubit–oscillator circuit beyond the ultrastrong-coupling regime},
  year         = {2016},
  issn         = {1745-2481},
  month        = oct,
  number       = {1},
  pages        = {44--47},
  volume       = {13},
  creationdate = {2025-08-20T16:42:26},
  doi          = {10.1038/nphys3906},
  fjournal     = {Nature Physics},
  publisher    = {Springer Science and Business Media LLC},
}

@Article{Niemczyk2010,
  author       = {Niemczyk, T. and Deppe, F. and Huebl, H. and Menzel, E. P. and Hocke, F. and Schwarz, M. J. and Garcia-Ripoll, J. J. and Zueco, D. and Hümmer, T. and Solano, E. and Marx, A. and Gross, R.},
  journal      = {Nat. Phys.},
  title        = {Circuit quantum electrodynamics in the ultrastrong-coupling regime},
  year         = {2010},
  issn         = {1745-2481},
  month        = jul,
  number       = {10},
  pages        = {772--776},
  volume       = {6},
  creationdate = {2025-08-20T16:32:29},
  doi          = {10.1038/nphys1730},
  fjournal     = {Nature Physics},
  publisher    = {Springer Science and Business Media LLC},
}

@Article{FornDiaz2019,
  author       = {Forn-D\'{\i}az, P. and Lamata, L. and Rico, E. and Kono, J. and Solano, E.},
  journal      = {Rev. Mod. Phys.},
  title        = {Ultrastrong coupling regimes of light-matter interaction},
  year         = {2019},
  month        = {Jun},
  pages        = {025005},
  volume       = {91},
  creationdate = {2025-08-20T16:56:53},
  doi          = {10.1103/RevModPhys.91.025005},
  issue        = {2},
  numpages     = {48},
  publisher    = {American Physical Society},
  url          = {https://link.aps.org/doi/10.1103/RevModPhys.91.025005},
}

@Article{Brune1996,
  author       = {Brune, M. and Schmidt-Kaler, F. and Maali, A. and Dreyer, J. and Hagley, E. and Raimond, J. M. and Haroche, S.},
  journal      = {Phys. Rev. Lett.},
  title        = {Quantum Rabi Oscillation: A Direct Test of Field Quantization in a Cavity},
  year         = {1996},
  month        = {Mar},
  pages        = {1800--1803},
  volume       = {76},
  creationdate = {2025-08-20T18:31:09},
  doi          = {10.1103/PhysRevLett.76.1800},
  issue        = {11},
  numpages     = {0},
  publisher    = {American Physical Society},
  url          = {https://link.aps.org/doi/10.1103/PhysRevLett.76.1800},
}

@Article{Zagoskin2008,
  author       = {Zagoskin, A. M. and Il'ichev, E. and McCutcheon, M. W. and Young, Jeff F. and Nori, Franco},
  journal      = {Phys. Rev. Lett.},
  title        = {Controlled Generation of Squeezed States of Microwave Radiation in a Superconducting Resonant Circuit},
  year         = {2008},
  month        = {Dec},
  pages        = {253602},
  volume       = {101},
  creationdate = {2025-08-20T18:37:04},
  doi          = {10.1103/PhysRevLett.101.253602},
  issue        = {25},
  numpages     = {4},
  publisher    = {American Physical Society},
  url          = {https://link.aps.org/doi/10.1103/PhysRevLett.101.253602},
}

@Article{Ashhab2010,
  author       = {Ashhab, S. and Nori, Franco},
  journal      = {Phys. Rev. A},
  title        = {Qubit-oscillator systems in the ultrastrong-coupling regime and their potential for preparing nonclassical states},
  year         = {2010},
  month        = {Apr},
  pages        = {042311},
  volume       = {81},
  creationdate = {2025-08-20T21:04:33},
  doi          = {10.1103/PhysRevA.81.042311},
  issue        = {4},
  numpages     = {17},
  publisher    = {American Physical Society},
  url          = {https://link.aps.org/doi/10.1103/PhysRevA.81.042311},
}

@Article{Lo2015,
  author       = {Lo, Hsiang-Yu and Kienzler, Daniel and de Clercq, Ludwig and Marinelli, Matteo and Negnevitsky, Vlad and Keitch, Ben C. and Home, Jonathan P.},
  journal      = {Nature},
  title        = {Spin–motion entanglement and state diagnosis with squeezed oscillator wavepackets},
  year         = {2015},
  issn         = {1476-4687},
  month        = may,
  number       = {7552},
  pages        = {336--339},
  volume       = {521},
  creationdate = {2025-08-21T00:20:07},
  doi          = {10.1038/nature14458},
  publisher    = {Springer Science and Business Media LLC},
}

@Article{Ashhab2013,
  author       = {Ashhab, S.},
  journal      = {Phys. Rev. A},
  title        = {Superradiance transition in a system with a single qubit and a single oscillator},
  year         = {2013},
  month        = {Jan},
  pages        = {013826},
  volume       = {87},
  creationdate = {2025-08-21T00:39:48},
  doi          = {10.1103/PhysRevA.87.013826},
  issue        = {1},
  numpages     = {6},
  publisher    = {American Physical Society},
  url          = {https://link.aps.org/doi/10.1103/PhysRevA.87.013826},
}

@Article{Cai2021,
  author       = {Cai, M.-L. and Liu, Z.-D. and Zhao, W.-D. and Wu, Y.-K. and Mei, Q.-X. and Jiang, Y. and He, L. and Zhang, X. and Zhou, Z.-C. and Duan, L.-M.},
  journal      = {Nat. Commun.},
  title        = {Observation of a quantum phase transition in the quantum Rabi model with a single trapped ion},
  year         = {2021},
  issn         = {2041-1723},
  month        = feb,
  number       = {1},
  pages        = {1126},
  volume       = {12},
  creationdate = {2025-08-22T11:47:16},
  day          = {18},
  doi          = {10.1038/s41467-021-21425-8},
  fjournal     = {Nature Communications},
  publisher    = {Springer Science and Business Media LLC},
  url          = {https://doi.org/10.1038/s41467-021-21425-8},
}

@Book{Sachdev2011,
  author       = {Sachdev, Subir},
  publisher    = {Cambridge University Press},
  title        = {Quantum {P}hase {T}ransitions},
  year         = {2011},
  edition      = {2},
  creationdate = {2025-08-22T01:01:04},
  place        = {Cambridge},
}

@Article{Zhu2020,
  author       = {Zhu, Han-Jie and Xu, Kai and Zhang, Guo-Feng and Liu, Wu-Ming},
  journal      = {Phys. Rev. Lett.},
  title        = {Finite-Component Multicriticality at the Superradiant Quantum Phase Transition},
  year         = {2020},
  month        = {Jul},
  pages        = {050402},
  volume       = {125},
  creationdate = {2025-08-22T01:27:37},
  doi          = {10.1103/PhysRevLett.125.050402},
  issue        = {5},
  numpages     = {6},
  publisher    = {American Physical Society},
  url          = {https://link.aps.org/doi/10.1103/PhysRevLett.125.050402},
}

@Article{Beaulieu2025,
  author       = {Beaulieu, Guillaume and Minganti, Fabrizio and Frasca, Simone and Scigliuzzo, Marco and Felicetti, Simone and Di Candia, Roberto and Scarlino, Pasquale},
  journal      = {PRX Quantum},
  title        = {Criticality-Enhanced Quantum Sensing with a Parametric Superconducting Resonator},
  year         = {2025},
  month        = {Apr},
  pages        = {020301},
  volume       = {6},
  creationdate = {2025-08-22T01:29:13},
  doi          = {10.1103/PRXQuantum.6.020301},
  issue        = {2},
  numpages     = {16},
  publisher    = {American Physical Society},
  url          = {https://link.aps.org/doi/10.1103/PRXQuantum.6.020301},
}

@Article{Puebla2017,
  author       = {Puebla, Ricardo and Hwang, Myung-Joong and Casanova, Jorge and Plenio, Martin B.},
  journal      = {Phys. Rev. Lett.},
  title        = {Probing the Dynamics of a Superradiant Quantum Phase Transition with a Single Trapped Ion},
  year         = {2017},
  month        = {Feb},
  pages        = {073001},
  volume       = {118},
  creationdate = {2025-08-22T01:30:15},
  doi          = {10.1103/PhysRevLett.118.073001},
  issue        = {7},
  numpages     = {5},
  publisher    = {American Physical Society},
  url          = {https://link.aps.org/doi/10.1103/PhysRevLett.118.073001},
}

@Article{Garbe2020,
  author       = {Garbe, Louis and Bina, Matteo and Keller, Arne and Paris, Matteo G. A. and Felicetti, Simone},
  journal      = {Phys. Rev. Lett.},
  title        = {Critical Quantum Metrology with a Finite-Component Quantum Phase Transition},
  year         = {2020},
  month        = {Mar},
  pages        = {120504},
  volume       = {124},
  creationdate = {2025-08-22T01:33:38},
  doi          = {10.1103/PhysRevLett.124.120504},
  issue        = {12},
  numpages     = {5},
  publisher    = {American Physical Society},
  url          = {https://link.aps.org/doi/10.1103/PhysRevLett.124.120504},
}

@Article{Chu2021,
  author       = {Chu, Yaoming and Zhang, Shaoliang and Yu, Baiyi and Cai, Jianming},
  journal      = {Phys. Rev. Lett.},
  title        = {Dynamic Framework for Criticality-Enhanced Quantum Sensing},
  year         = {2021},
  month        = {Jan},
  pages        = {010502},
  volume       = {126},
  creationdate = {2025-08-22T01:36:09},
  doi          = {10.1103/PhysRevLett.126.010502},
  issue        = {1},
  numpages     = {7},
  publisher    = {American Physical Society},
  url          = {https://link.aps.org/doi/10.1103/PhysRevLett.126.010502},
}

@Article{Ilias2022,
  author       = {Ilias, Theodoros and Yang, Dayou and Huelga, Susana F. and Plenio, Martin B.},
  journal      = {PRX Quantum},
  title        = {Criticality-Enhanced Quantum Sensing via Continuous Measurement},
  year         = {2022},
  month        = {Mar},
  pages        = {010354},
  volume       = {3},
  creationdate = {2025-08-18T01:40:07},
  doi          = {10.1103/PRXQuantum.3.010354},
  issue        = {1},
  numpages     = {21},
  publisher    = {American Physical Society},
  url          = {https://link.aps.org/doi/10.1103/PRXQuantum.3.010354},
}

@Article{Yang2023,
  author       = {Yang, Dayou and Huelga, Susana F. and Plenio, Martin B.},
  journal      = {Phys. Rev. X},
  title        = {Efficient Information Retrieval for Sensing via Continuous Measurement},
  year         = {2023},
  month        = {Jul},
  pages        = {031012},
  volume       = {13},
  creationdate = {2025-08-22T01:37:51},
  doi          = {10.1103/PhysRevX.13.031012},
  issue        = {3},
  numpages     = {27},
  publisher    = {American Physical Society},
  url          = {https://link.aps.org/doi/10.1103/PhysRevX.13.031012},
}

@Article{Felicetti2020,
  author       = {Felicetti, Simone and Le Boit\'e, Alexandre},
  journal      = {Phys. Rev. Lett.},
  title        = {Universal Spectral Features of Ultrastrongly Coupled Systems},
  year         = {2020},
  month        = {Jan},
  pages        = {040404},
  volume       = {124},
  creationdate = {2025-08-22T01:39:40},
  doi          = {10.1103/PhysRevLett.124.040404},
  issue        = {4},
  numpages     = {6},
  publisher    = {American Physical Society},
  url          = {https://link.aps.org/doi/10.1103/PhysRevLett.124.040404},
}

@Article{Lv2018,
  author       = {Lv, Dingshun and An, Shuoming and Liu, Zhenyu and Zhang, Jing-Ning and Pedernales, Julen S. and Lamata, Lucas and Solano, Enrique and Kim, Kihwan},
  journal      = {Phys. Rev. X},
  title        = {Quantum Simulation of the Quantum Rabi Model in a Trapped Ion},
  year         = {2018},
  month        = {Apr},
  pages        = {021027},
  volume       = {8},
  creationdate = {2025-08-18T01:46:15},
  doi          = {10.1103/PhysRevX.8.021027},
  issue        = {2},
  numpages     = {11},
  publisher    = {American Physical Society},
  url          = {https://link.aps.org/doi/10.1103/PhysRevX.8.021027},
}

@Article{Chen2021,
  author       = {Chen, Xi and Wu, Ze and Jiang, Min and Lü, Xin-You and Peng, Xinhua and Du, Jiangfeng},
  journal      = {Nat. Commun.},
  title        = {Experimental quantum simulation of superradiant phase transition beyond no-go theorem via antisqueezing},
  year         = {2021},
  issn         = {2041-1723},
  month        = nov,
  number       = {1},
  pages        = {6281},
  volume       = {12},
  creationdate = {2025-08-22T11:49:14},
  day          = {01},
  doi          = {10.1038/s41467-021-26573-5},
  fjournal     = {Nature Communications},
  publisher    = {Springer Science and Business Media LLC},
  url          = {https://doi.org/10.1038/s41467-021-26573-5},
}

@Article{Wu2024,
  author       = {Wu, Ze and Hu, Changsheng and Wang, Tianyun and Chen, Yuquan and Li, Yuchen and Zhao, Liqiang and L\"u, Xin-You and Peng, Xinhua},
  journal      = {Phys. Rev. Lett.},
  title        = {Experimental Quantum Simulation of Multicriticality in Closed and Open Rabi Model},
  year         = {2024},
  month        = {Oct},
  pages        = {173602},
  volume       = {133},
  creationdate = {2025-08-18T01:46:26},
  doi          = {10.1103/PhysRevLett.133.173602},
  issue        = {17},
  numpages     = {6},
  publisher    = {American Physical Society},
  url          = {https://link.aps.org/doi/10.1103/PhysRevLett.133.173602},
}

@Article{Braumueller2017,
  author       = {Braum{\"u}ller, Jochen and Marthaler, Michael and Schneider, Andre and Stehli, Alexander and Rotzinger, Hannes and Weides, Martin and Ustinov, Alexey V.},
  journal      = {Nat. Commun.},
  title        = {Analog quantum simulation of the Rabi model in the ultra-strong coupling regime},
  year         = {2017},
  issn         = {2041-1723},
  month        = {Oct},
  number       = {1},
  pages        = {779},
  volume       = {8},
  creationdate = {2025-08-22T02:03:17},
  day          = {03},
  doi          = {10.1038/s41467-017-00894-w},
  fjournal     = {Nature Communications},
  url          = {https://doi.org/10.1038/s41467-017-00894-w},
}

@Article{Langford2017,
  author       = {Langford, N. K. and Sagastizabal, R. and Kounalakis, M. and Dickel, C. and Bruno, A. and Luthi, F. and Thoen, D. J. and Endo, A. and DiCarlo, L.},
  journal      = {Nat. Commun.},
  title        = {Experimentally simulating the dynamics of quantum light and matter at deep-strong coupling},
  year         = {2017},
  issn         = {2041-1723},
  month        = nov,
  number       = {1},
  pages        = {1715},
  volume       = {8},
  creationdate = {2025-08-22T11:45:35},
  day          = {23},
  doi          = {10.1038/s41467-017-01061-x},
  fjournal     = {Nature Communications},
  publisher    = {Springer Science and Business Media LLC},
  url          = {https://doi.org/10.1038/s41467-017-01061-x},
}

@Article{Lyu2024,
  author       = {Lyu, Guitao and Kottmann, Korbinian and Plenio, Martin B. and Hwang, Myung-Joong},
  journal      = {Phys. Rev. Res.},
  title        = {Multicritical dissipative phase transitions in the anisotropic open quantum Rabi model},
  year         = {2024},
  month        = {Jul},
  pages        = {033075},
  volume       = {6},
  creationdate = {2025-08-18T01:46:01},
  doi          = {10.1103/PhysRevResearch.6.033075},
  issue        = {3},
  numpages     = {14},
  publisher    = {American Physical Society},
  url          = {https://link.aps.org/doi/10.1103/PhysRevResearch.6.033075},
}

@Article{Li2024,
  author       = {Li, Jiahui and Fazio, Rosario and Wang, Yingdan and Chesi, Stefano},
  journal      = {Phys. Rev. Res.},
  title        = {Spin fluctuations in the dissipative phase transitions of the quantum Rabi model},
  year         = {2024},
  month        = {Dec},
  pages        = {043250},
  volume       = {6},
  creationdate = {2025-08-22T02:16:05},
  doi          = {10.1103/PhysRevResearch.6.043250},
  issue        = {4},
  numpages     = {18},
  publisher    = {American Physical Society},
  url          = {https://link.aps.org/doi/10.1103/PhysRevResearch.6.043250},
}

@Article{Gerry2024,
  author       = {Gerry, Matthew and Kewming, Michael J. and Segal, Dvira},
  journal      = {Phys. Rev. Res.},
  title        = {Understanding multiple timescales in quantum dissipative dynamics: Insights from quantum trajectories},
  year         = {2024},
  month        = {Jul},
  pages        = {033106},
  volume       = {6},
  creationdate = {2025-08-19T00:58:35},
  doi          = {10.1103/PhysRevResearch.6.033106},
  issue        = {3},
  numpages     = {15},
  publisher    = {American Physical Society},
  url          = {https://link.aps.org/doi/10.1103/PhysRevResearch.6.033106},
}

@Article{Lee2012,
  author       = {Lee, Tony E. and H\"affner, H. and Cross, M. C.},
  journal      = {Phys. Rev. Lett.},
  title        = {Collective Quantum Jumps of Rydberg Atoms},
  year         = {2012},
  month        = {Jan},
  pages        = {023602},
  volume       = {108},
  creationdate = {2025-08-23T17:52:46},
  doi          = {10.1103/PhysRevLett.108.023602},
  issue        = {2},
  numpages     = {5},
  publisher    = {American Physical Society},
  url          = {https://link.aps.org/doi/10.1103/PhysRevLett.108.023602},
}

@Article{Ates2012,
  author       = {Ates, Cenap and Olmos, Beatriz and Garrahan, Juan P. and Lesanovsky, Igor},
  journal      = {Phys. Rev. A},
  title        = {Dynamical phases and intermittency of the dissipative quantum Ising model},
  year         = {2012},
  month        = {Apr},
  pages        = {043620},
  volume       = {85},
  creationdate = {2025-08-18T21:24:48},
  doi          = {10.1103/PhysRevA.85.043620},
  issue        = {4},
  numpages     = {8},
  publisher    = {American Physical Society},
  url          = {https://link.aps.org/doi/10.1103/PhysRevA.85.043620},
}

@Article{Lesanovsky2013,
  author       = {Lesanovsky, Igor and van Horssen, Merlijn and Guta, Madalin and Garrahan, Juan P.},
  journal      = {Phys. Rev. Lett.},
  title        = {Characterization of Dynamical Phase Transitions in Quantum Jump Trajectories Beyond the Properties of the Stationary State},
  year         = {2013},
  month        = {Apr},
  pages        = {150401},
  volume       = {110},
  creationdate = {2025-08-29T13:09:15},
  doi          = {10.1103/PhysRevLett.110.150401},
  issue        = {15},
  numpages     = {5},
  publisher    = {American Physical Society},
  url          = {https://link.aps.org/doi/10.1103/PhysRevLett.110.150401},
}

@Article{Letscher2017,
  author       = {Letscher, F. and Thomas, O. and Niederpr\"um, T. and Fleischhauer, M. and Ott, H.},
  journal      = {Phys. Rev. X},
  title        = {Bistability Versus Metastability in Driven Dissipative Rydberg Gases},
  year         = {2017},
  month        = {May},
  pages        = {021020},
  volume       = {7},
  creationdate = {2025-08-28T22:23:12},
  doi          = {10.1103/PhysRevX.7.021020},
  issue        = {2},
  numpages     = {16},
  publisher    = {American Physical Society},
  url          = {https://link.aps.org/doi/10.1103/PhysRevX.7.021020},
}

@Article{Sibalic2016,
  author       = {Sibalic, N. and Wade, C. G. and Adams, C. S. and Weatherill, K. J. and Pohl, T.},
  journal      = {Phys. Rev. A},
  title        = {Driven-dissipative many-body systems with mixed power-law interactions: Bistabilities and temperature-driven nonequilibrium phase transitions},
  year         = {2016},
  month        = {Jul},
  pages        = {011401},
  volume       = {94},
  creationdate = {2025-08-29T13:18:40},
  doi          = {10.1103/PhysRevA.94.011401},
  issue        = {1},
  numpages     = {6},
  publisher    = {American Physical Society},
  url          = {https://link.aps.org/doi/10.1103/PhysRevA.94.011401},
}

@Article{Jin2024,
  author       = {Jin, Yuan-De and Qiu, Chu-Dan and Ma, Wen-Long},
  journal      = {Phys. Rev. A},
  title        = {Theory of metastability in discrete-time open quantum dynamics},
  year         = {2024},
  month        = {Apr},
  pages        = {042204},
  volume       = {109},
  creationdate = {2025-08-19T10:19:52},
  doi          = {10.1103/PhysRevA.109.042204},
  issue        = {4},
  numpages     = {13},
  publisher    = {American Physical Society},
  url          = {https://link.aps.org/doi/10.1103/PhysRevA.109.042204},
}

@Article{Malossi2014,
  author       = {Malossi, N. and Valado, M. M. and Scotto, S. and Huillery, P. and Pillet, P. and Ciampini, D. and Arimondo, E. and Morsch, O.},
  journal      = {Phys. Rev. Lett.},
  title        = {Full Counting Statistics and Phase Diagram of a Dissipative Rydberg Gas},
  year         = {2014},
  month        = {Jul},
  pages        = {023006},
  volume       = {113},
  creationdate = {2025-08-29T13:32:31},
  doi          = {10.1103/PhysRevLett.113.023006},
  issue        = {2},
  numpages     = {5},
  publisher    = {American Physical Society},
  url          = {https://link.aps.org/doi/10.1103/PhysRevLett.113.023006},
}

@Article{MendozaArenas2016,
  author       = {Mendoza-Arenas, J. J. and Clark, S. R. and Felicetti, S. and Romero, G. and Solano, E. and Angelakis, D. G. and Jaksch, D.},
  journal      = {Phys. Rev. A},
  title        = {Beyond mean-field bistability in driven-dissipative lattices: Bunching-antibunching transition and quantum simulation},
  year         = {2016},
  month        = {Feb},
  pages        = {023821},
  volume       = {93},
  creationdate = {2025-08-29T13:49:46},
  doi          = {10.1103/PhysRevA.93.023821},
  issue        = {2},
  numpages     = {11},
  publisher    = {American Physical Society},
  url          = {https://link.aps.org/doi/10.1103/PhysRevA.93.023821},
}

@Article{Melo2016,
  author       = {de Melo, Natalia R. and Wade, Christopher G. and \ifmmode \check{S}\else \v{S}\fi{}ibali\ifmmode \acute{c}\else \'{c}\fi{}, Nikola and Kondo, Jorge M. and Adams, Charles S. and Weatherill, Kevin J.},
  journal      = {Phys. Rev. A},
  title        = {Intrinsic optical bistability in a strongly driven Rydberg ensemble},
  year         = {2016},
  month        = {Jun},
  pages        = {063863},
  volume       = {93},
  creationdate = {2025-08-29T13:50:44},
  doi          = {10.1103/PhysRevA.93.063863},
  issue        = {6},
  numpages     = {5},
  publisher    = {American Physical Society},
  url          = {https://link.aps.org/doi/10.1103/PhysRevA.93.063863},
}

@Article{Alaeian2022,
  author       = {Alaeian, Hadiseh and Buča, Berislav},
  journal      = {Commun. Phys.},
  title        = {Exact multistability and dissipative time crystals in interacting fermionic lattices},
  year         = {2022},
  issn         = {2399-3650},
  month        = dec,
  pages        = {318},
  volume       = {5},
  creationdate = {2025-08-28T19:31:34},
  doi          = {10.1038/s42005-022-01090-z},
  publisher    = {Springer Science and Business Media LLC},
}

@Article{Ferri2021,
  author       = {Ferri, Francesco and Rosa-Medina, Rodrigo and Finger, Fabian and Dogra, Nishant and Soriente, Matteo and Zilberberg, Oded and Donner, Tobias and Esslinger, Tilman},
  journal      = {Phys. Rev. X},
  title        = {Emerging Dissipative Phases in a Superradiant Quantum Gas with Tunable Decay},
  year         = {2021},
  month        = {Dec},
  pages        = {041046},
  volume       = {11},
  creationdate = {2025-08-29T13:55:07},
  doi          = {10.1103/PhysRevX.11.041046},
  issue        = {4},
  numpages     = {20},
  publisher    = {American Physical Society},
  url          = {https://link.aps.org/doi/10.1103/PhysRevX.11.041046},
}

@Article{LeBoite2017,
  author       = {Le Boit\'e, Alexandre and Hwang, Myung-Joong and Plenio, Martin B.},
  journal      = {Phys. Rev. A},
  title        = {Metastability in the driven-dissipative Rabi model},
  year         = {2017},
  month        = {Feb},
  pages        = {023829},
  volume       = {95},
  creationdate = {2025-08-19T10:22:27},
  doi          = {10.1103/PhysRevA.95.023829},
  issue        = {2},
  numpages     = {11},
  publisher    = {American Physical Society},
  url          = {https://link.aps.org/doi/10.1103/PhysRevA.95.023829},
}

@Article{Hannukainen2018,
  author       = {Hannukainen, Julia and Larson, Jonas},
  journal      = {Phys. Rev. A},
  title        = {Dissipation-driven quantum phase transitions and symmetry breaking},
  year         = {2018},
  month        = {Oct},
  pages        = {042113},
  volume       = {98},
  creationdate = {2025-08-18T19:27:20},
  doi          = {10.1103/PhysRevA.98.042113},
  issue        = {4},
  numpages     = {11},
  publisher    = {American Physical Society},
  url          = {https://link.aps.org/doi/10.1103/PhysRevA.98.042113},
}

@Article{Soriente2018,
  author       = {Soriente, M. and Donner, T. and Chitra, R. and Zilberberg, O.},
  journal      = {Phys. Rev. Lett.},
  title        = {Dissipation-Induced Anomalous Multicritical Phenomena},
  year         = {2018},
  month        = {May},
  pages        = {183603},
  volume       = {120},
  creationdate = {2025-08-29T13:56:06},
  doi          = {10.1103/PhysRevLett.120.183603},
  issue        = {18},
  numpages     = {5},
  publisher    = {American Physical Society},
  url          = {https://link.aps.org/doi/10.1103/PhysRevLett.120.183603},
}

@Article{Minganti2021,
  author       = {Minganti, Fabrizio and Arkhipov, Ievgen I and Miranowicz, Adam and Nori, Franco},
  journal      = {New J. Phys.},
  title        = {Continuous dissipative phase transitions with or without symmetry breaking},
  year         = {2021},
  month        = {dec},
  number       = {12},
  pages        = {122001},
  volume       = {23},
  creationdate = {2025-08-22T01:32:06},
  doi          = {10.1088/1367-2630/ac3db8},
  fjournal     = {New Journal of Physics},
  publisher    = {IOP Publishing},
  url          = {https://dx.doi.org/10.1088/1367-2630/ac3db8},
}

@Article{Hedges2009,
  author       = {Lester O. Hedges and Robert L. Jack and Juan P. Garrahan and David Chandler},
  journal      = {Science},
  title        = {Dynamic Order-Disorder in Atomistic Models of Structural Glass Formers},
  year         = {2009},
  number       = {5919},
  pages        = {1309-1313},
  volume       = {323},
  creationdate = {2025-08-23T18:11:40},
  doi          = {10.1126/science.1166665},
  url          = {https://www.science.org/doi/abs/10.1126/science.1166665},
}

@Article{Rakovszky2024,
  author       = {Rakovszky, Tibor and Gopalakrishnan, Sarang and von Keyserlingk, Curt},
  journal      = {Phys. Rev. X},
  title        = {Defining Stable Phases of Open Quantum Systems},
  year         = {2024},
  month        = {Nov},
  pages        = {041031},
  volume       = {14},
  creationdate = {2025-08-28T22:08:16},
  doi          = {10.1103/PhysRevX.14.041031},
  issue        = {4},
  numpages     = {35},
  publisher    = {American Physical Society},
  url          = {https://link.aps.org/doi/10.1103/PhysRevX.14.041031},
}

@Article{Weimer2015,
  author       = {Weimer, Hendrik},
  journal      = {Phys. Rev. Lett.},
  title        = {Variational Principle for Steady States of Dissipative Quantum Many-Body Systems},
  year         = {2015},
  month        = {Jan},
  pages        = {040402},
  volume       = {114},
  creationdate = {2025-08-29T14:59:06},
  doi          = {10.1103/PhysRevLett.114.040402},
  issue        = {4},
  numpages     = {6},
  publisher    = {American Physical Society},
  url          = {https://link.aps.org/doi/10.1103/PhysRevLett.114.040402},
}

@Article{Weimer2015a,
  author       = {Weimer, Hendrik},
  journal      = {Phys. Rev. A},
  title        = {Variational analysis of driven-dissipative Rydberg gases},
  year         = {2015},
  month        = {Jun},
  pages        = {063401},
  volume       = {91},
  creationdate = {2025-08-29T15:00:00},
  doi          = {10.1103/PhysRevA.91.063401},
  issue        = {6},
  numpages     = {7},
  publisher    = {American Physical Society},
  url          = {https://link.aps.org/doi/10.1103/PhysRevA.91.063401},
}

@Article{Gangat2017,
  author       = {Gangat, Adil A. and I, Te and Kao, Ying-Jer},
  journal      = {Phys. Rev. Lett.},
  title        = {Steady States of Infinite-Size Dissipative Quantum Chains via Imaginary Time Evolution},
  year         = {2017},
  month        = {Jul},
  pages        = {010501},
  volume       = {119},
  creationdate = {2025-08-29T15:00:46},
  doi          = {10.1103/PhysRevLett.119.010501},
  issue        = {1},
  numpages     = {5},
  publisher    = {American Physical Society},
  url          = {https://link.aps.org/doi/10.1103/PhysRevLett.119.010501},
}

\end{document}